\documentclass{aa}  
\usepackage{natbib}

\usepackage{color}

\usepackage{graphicx}
\usepackage{txfonts}
\usepackage[switch]{lineno} 
\begin{document}
\title{Reassessing the Spin of Second-born Black Holes in Coalescing Binary Black Holes and Its Connection to the $\chi_{\rm eff}$ – $q$ Correlation}
\titlerunning{Reassessing the spin of the second-born BH and the $\chi_{\rm eff}$ – $q$ correlation}

\author{Zi-Yuan Wang\inst{1} 
         \and
         Ying Qin\inst{1}   
         \and 
          Rui-Chong Hu\inst{2,3}
        \and 
         Yuan-Zhu Wang\inst{4}  
        \and
        Georges Meynet\inst{5,6}
        \and
        Han-Feng Song\inst{7}
        }
\authorrunning{Wang et al.}
\institute{Department of Physics, Anhui Normal University, Wuhu, Anhui, 241002, China \\
        \email{yingqin@ahnu.edu.cn}        
        \and 
           Nevada Center for Astrophysics, University of Nevada, Las Vegas, NV 89154, USA
        \and
           Department of Physics and Astronomy, University of Nevada, Las Vegas, NV 89154, USA
        \and 
            Institute for Theoretical Physics and Cosmology, Zhejiang University of Technology, Hangzhou, 310032, China   
        \and 
            Département d’Astronomie, Université de Genève, Chemin Pegasi 51, 1290 Versoix, Switzerland
        \and 
            Gravitational Wave Science Center (GWSC), Université de Genève, 24 quai E. Ansermet, 1211 Geneva, Switzerland
        \and     
           College of Physics, Guizhou University, Guiyang city, Guizhou Province, 550025, China}


 \abstract
   {The mass ratio $q$ and effective inspiral spin $\chi_{\rm eff}$ of binary black hole mergers in GWTC-4.0 have been reported to display a weaker anti-correlation compared to GWTC-3.0, a feature whose origin has been explored by several groups. In this work, within the isolated binary evolution framework, we adopt a recently proposed wind prescription for helium stars to systematically investigate the spin of the second-born black hole and its role in shaping this correlation.}
   {Our first goal is to investigate the main factors shaping the spin of the second-born black hole in a helium star–black hole binary, whether formed via a common-envelope or stable mass-transfer channel, and to further explore the potential correlation between the mass ratio $q$ and the effective inspiral spin $\chi_{\rm eff}$.}
   {Using the stellar and binary evolution code \texttt{MESA}, which includes a recently proposed helium-star wind prescription alongside internal differential rotation and tidal interactions, we investigate how initial conditions and fundamental physical processes shape the spin of the resulting black hole. Additionally, we further employ rapid population synthesis calculations with \texttt{COMPAS} to predict the correlation between the mass ratio $q$ and the effective inspiral spin $\chi_{\rm eff}$.}
   {We find that the recently proposed wind prescription for helium stars is substantially weaker than the standard Dutch wind scheme, particularly at subsolar metallicity. Using this scheme, we perform detailed binary modeling of a helium star with a black hole companion. Our results show that the spin magnitude of the resulting black hole is insensitive to the helium star’s evolutionary stage at the onset of tidal interactions or to the companion mass. Instead, wind mass loss plays the dominant role: more massive helium-star progenitors produce lower-spinning black holes. The initial stellar rotation has only a minor effect, especially under strong tidal coupling, consistent with the common assumption of orbital synchronization. We then provide a fitting formula for the spin magnitude of the resulting second-born black hole. By contrast, the efficiency of angular momentum transport within helium stars can significantly alter the spin magnitude of the resulting black hole.}
   {Combining the fitting formula provided from the detailed binary evolution and rapid population synthesis with default model assumptions, we find that in the stable mass-transfer channel the majority (85.8\%) of binary black holes undergo mass-ratio reversal, whereas in the common-envelope channel, only a small fraction (2.8\%) exhibit mass-ratio reversal. Notably, we find no correlation between the mass ratio $q$ and the effective spin parameter $\chi_{\rm eff}$ in either evolutionary channel. In future work, we plan to investigate how alternative physical prescriptions in population-synthesis models influence the relationship between $q$ and $\chi_{\rm eff}$, and to compare our predictions with coalescing binary black holes reported by the LIGO–Virgo–KAGRA Collaboration.}


\maketitle

\section{Introduction}\label{sect1}
By the end of the third observing run (O3), the LIGO–Virgo–KAGRA (LVK) collaboration \citep{acernese2015,Akutsu2021} had reported a catalog of 69 gravitational-wave (GW) candidates with a false-alarm rate below $1~\mathrm{yr}^{-1}$, referred to as GWTC-3.0 \citep{GWTC-3}. More recently, the LVK collaboration released GWTC-4.0 \citep{GWTC4_catlog}, which includes 84 additional binary black hole (BBH) candidates with the same false-alarm threshold. This growing BBH population has enabled more detailed studies of the distributions of black hole mass, spin, and redshift \citep{Roulet2020,Abbott2021}, offering valuable insights into the mechanisms that govern binary black hole formation.

In general, the effective inspiral spin has been widely recognized as a potential probe for distinguishing BBH formation channels. Beyond one-dimensional distributions, correlations between key parameters provide crucial insights into BBH formation scenarios. Using BBH events from the second GW transient catalog (GWTC-2) \citep{GTWC2}, \citet{Callister2021} first identified an anti-correlation between the mass ratio $q$ and the effective inspiral spin $\chi_{\rm eff}$. \citet{Adamcewicz2022} later reinforced this finding through a statistical copula analysis, showing that BBHs in GWTC-2 with unequal component masses tend to have higher $\chi_{\rm eff}$ (98.7\% credibility). This correlation was further validated with BBH events in GWTC-3.0 \citep{GWTC-3-pop}. Using updated GWTC-3.0 data, \citet{Adamcewicz2023} improved their copula-based framework and found evidence for an anti-correlation between $q$ and $\chi_{\rm eff}$ with 99.7\% credibility. While preparing this work, we noted that the BBH sample was expanded with new candidates reported in GWTC-4.0 \citep{GWTC4_catlog}. An updated population study by \cite{GWTC4_pop} found that $q$ and $\chi_{\rm eff}$ are anti-correlated at 92\% credibility. 

Various studies have explored the origin of the $q-\chi_{\rm eff}$ anti-correlation. One plausible explanation is that BBHs assembled in active galactic nucleus (AGN) disks naturally exhibit such a trend \citep[e.g.,][]{McKernan2022,Santini2023,Cook2024,Delfavero2024,Li2024,Li2025}. In the context of isolated binary evolution, \citet{Bavera2021} showed that, under the assumption of a high common-envelope ejection efficiency, binaries undergoing common-envelope evolution can also reproduce the observed anti-correlation. \citet{Zevin2022} found that stable mass transfer leading to mass ratio reversal may similarly account for the trend. \citet{Broekgaarden2022} systematically explored 560 population synthesis models and showed that the more massive BH forms second in $\gtrsim 70\%$ of BBHs observable by the LVK. In addition, \citet{Olejak2024} argued that a combination of common-envelope \citep[CE, e.g.,][]{Tutukov1973,Phinney1991,Ivanova2013,Belczynski2007,Belczynski2016} and stable mass transfer channels \citep[SMT, e.g.,][]{van2017,Monica2021} provides a good match to the observed correlation, while \citet{Banerjee2024} demonstrated that isolated massive binary evolution via the stable mass transfer subchannel can reproduce the key observed features of BBH populations, including their masses, mass ratios, and spins. More recently, \cite{Xu2025} demonstrated, using detailed binary-evolution models, that the SMT channel is a robust contributor to the observed BBH mergers. Notably, \citet{Klencki2025} found that the SMT may not produce very short-period BH–He star binaries due to a dynamical instability that sets in their models of short-period BH+OB star binaries.

In isolated binary evolution, the common-envelope channel has been extensively studied as a pathway to BBH mergers \cite[e.g.,][]{Belczynski2016,Eldridge2016,Stevenson2017,Qin2018,Kruckow2018,Spera2019,Mapelli2019,Marchant2019,Bavera2020}. Within this framework, \cite{Qin2018} showed that the spin of the first-born BH is typically negligible,\footnote{See also \cite{Fuller2019}; \cite{Belczynski2020} reported a small but nonzero value, $\chi \sim 0.1$.} whereas the second-born BH can span the full range (i.e., from nonspinning to maximally spinning). Its spin is primarily determined by the competition between stellar winds and tidal interactions during the helium-star (He-star) phase \citep{Qin2018}. However, the strength of He-star winds remains uncertain. Recently, \cite{Sander&Vink2020} proposed the first theoretically motivated prescription for mass loss in massive He stars, and \cite{Sander2023} further suggested incorporating a temperature dependence into the mass-loss rate. In parallel, \cite{Sciarini2024} identified inconsistencies in the implementation of dynamical tides relative to the original formulation of \cite{Zahn1977}. This issue was corrected and further tested by \cite{Qin2024_gap}, who found that dynamical tides are slightly weaker than previously assumed \citep[see details in][]{Qin2024_gap}. The combination of this new wind prescription with the revised tidal treatment enables us to investigate the spin of the BH formed from a He star in a close binary system. Moreover, we try to explore the relation between the mass ratio $q$ and the effective inspiral spin $\chi_{\rm eff}$ in binaries consisting of a He-star progenitor and a BH companion.

In this work, we employ detailed binary-evolution calculations, incorporating a recently proposed wind prescription for He stars and a revised implementation of dynamical tides, to investigate the key factors that determine the spin of the second-born BH formed from a He-star progenitor with a BH companion. We then derive a fitting formula for the spin magnitude of the resulting second-born BH. Using BH–He star populations generated by rapid population-synthesis calculations, we further investigate the correlation between the mass ratio $q$ and the effective inspiral spin $\chi_{\rm eff}$ in BBH populations. In Sect.~\ref{sect2}, we outline the main physical assumptions underlying our detailed binary evolution models and provide a brief introduction to the He-star wind mass loss. We then present our main findings in Sect.~\ref{sect3}. Finally, we summarize the conclusions with some discussion in Sect.~\ref{sect4}.

\section{Methods}\label{sect2}
\subsection{Main physics adopted in this work}

We performed detailed binary modeling using the release version \texttt{mesa-r15140} of the Modules for Experiments in Stellar Astrophysics (\texttt{MESA}) stellar evolution code \citep{Paxton2011, Paxton2013, Paxton2015, Paxton2018, Paxton2019, Jermyn2023}. Our He star models were constructed following the methodology outlined in recent studies \citep[e.g.,][]{Fragos2023,lv2023,Qin2024_gap,Qin2024_casebb}. Throughout this work, we adopted a solar metallicity of $Z_{\odot} = $ 0.0142 \citep{Asplund2009}.

We modeled convection using the mixing-length theory \citep{MLT1958}, adopting a mixing-length parameter of $\alpha_{\rm mlt}=1.93$, consistent with the value used in the \texttt{POSYDON} framework \citep{Fragos2023}. Convective boundaries were determined based on the Ledoux criterion, with a step-overshooting parameter of $\alpha_{\rm p} = 0.335\, H_{\rm p}$, calibrated to match the observed drop in rotation rates of massive main-sequence stars \citep{Brott2011}, where $H_{\rm p}$ represents the pressure scale height at the Ledoux boundary. Semiconvection was included in the He star models following \cite{Langer1983}, with an efficiency parameter of $\alpha_{\rm sc}=1.0$. For nucleosynthesis calculations, we employed the \texttt{approx21.net} reaction network.

We modeled rotational mixing and angular momentum transport as diffusive processes \citep{Heger2000}, incorporating the effects of the Goldreich–Schubert–Fricke instability, Eddington–Sweet circulations, as well as secular and dynamical shear mixing. The efficiency of diffusive element mixing was set to $f_{\rm c} = 1/30$, following \cite{Chaboyer1992,Heger2000}. To account for the sensitivity of the $\mu$-gradient to rotationally induced mixing, we mitigated its impact by multiplying $f_\mu = 0.05$, as recommended by \cite{Heger2000}.

Additionally, we accounted for rotationally enhanced mass loss following \cite{Heger1998} and \cite{Langer1998}:
\begin{equation}\label{ml}
    \centering
    \dot{M}(\omega)= \dot{M}(0)\left(\frac{1}{1-\omega/\omega_{\rm crit}}\right)^\xi,
\end{equation}

where $\omega$ is the angular velocity and $\omega_{\rm crit}$ is the critical angular velocity at the stellar surface. The latter is given by
\begin{equation}\label{ml}
    \centering
\omega_{\rm crit}^2 = (1- L/L_{\rm Edd})GM/R^3,
\end{equation}
where $L$, $M$, and $R$ are the stellar luminosity, mass, and radius, respectively, and $G$ is the gravitational constant. The classical Eddington luminosity, $L_{\rm Edd}$, is defined using the electron-scattering opacity for a fully ionized medium, $\kappa$ = 0.2(1 + X) $\rm cm^2$ $\rm g^{-1}$, with $\rm X$ = 0 for a hydrogen-free WR star envelope. We adopted an exponent of $\xi = 0.43$ \citep{Langer1998}. Notably, we did not include gravity-darkening effects, as discussed in \cite{Maeder2000}.

We applied the theory of dynamical tides to He star stars with radiative envelopes, following the framework of \cite{Zahn1977}. The synchronization timescale was computed using the prescriptions of \cite{Zahn1977}, \cite{Hut1981}, and \cite{Hurley2002}, while the tidal torque coefficient $E_2$ was adopted from the updated fitting formula of \cite{Qin2018}. Given previous inconsistencies in the implementation of the synchronization timescale \citep{Sciarini2024}, we adopted the corrected version from \cite{Qin2024_gap}. We adopted the Jeans mode mass loss, i.e., the stellar wind removes the specific angular momentum of the mass-losing star. In our binary modeling, we evolved He stars until their central carbon depletion is reached. Furthermore, we adopted a direct collapse scenario for BH formation (assuming no mass and angular momentum loss), implying that newly formed BHs do not experience mass loss or natal kicks \citep{Belczynski2008}.

\subsection{He star mass-loss rates}\label{set2.2}
He stars represent the final evolutionary stage prior to BH formation, making their mass loss a key determinant of the resulting BH mass \citep{Woosley2020}. Stellar evolution models have traditionally relied on empirically derived mass-loss prescriptions for He stars. Based on the mass-loss rate of \citet{Yoon2017}, \citet{Woosley2019} investigated the evolution of massive He stars. Notably, the recent work by \citet{Sander&Vink2020} introduced the first theoretically grounded mass-loss formulae for massive He stars. \citet{Higgins2021} adopted this prescription to compare with earlier empirical recipes widely used in stellar evolution and population-synthesis modelling. This wind prescription provides a physically motivated description of mass loss, derived from dynamically consistent atmosphere models for He stars by \citet{Sander&Vink2020} (hereafter SV2020), namely,

\begin{equation}\label{}
    \centering
    \log \left(\frac{\dot{M}_{\rm SV2020}}{M_\odot\,\rm yr^{-1}}\right) = \alpha \log(\log L - \log L_0)+\frac{3}{4}\log \frac{L}{10L_0}+\log \dot{M}_{10},
\end{equation}
where
\begin{equation}\label{}
    \centering
    \alpha = 0.32 \log Z_{\rm i}+1.40,
\end{equation}
\begin{equation}\label{L_L0}
    \centering
    \log L_0/L_{\odot} = -0.87\log Z_{\rm i}+5.06,
\end{equation}
\begin{equation}\label{}
    \centering
    \log \left(\frac{\dot{M}_{10}}{M_\odot\,\rm yr^{-1}}\right) =-0.75 \log Z_{\rm i}-4.06.
\end{equation}
In the above equations, $Z_{\rm i}$ is the initial metallicity (in units of ${Z_\odot}$) and $L_{\rm 0}$ represents the asymptotic limit for which there is theoretically zero mass-loss. The exponent $\alpha$ characterizes the curvature of the breakdown, and $\dot{M}_{\rm 10}$ denotes the mass-loss rate at $L = 10\,L_{\rm 0}$. When $L \leq L_{\rm 0}$,  the luminosity is insufficient to support the optically thick winds, leading to the onset of the “breakdown regime'', where the mass loss becomes zero. To account for this, we adopted the wind prescription proposed by \cite{Vink2017} (hereafter V2017) as the lower limit, namely,
\begin{equation}\label{}
    \centering
     \log \left(\frac{\dot{M}_{\rm V2017}}{M_\odot\,\rm yr^{-1}}\right) = -13.3 + 1.36 \log\, (L/L_{\rm \odot}) + 0.61 \log\,(Z_{\rm cur}/Z_{\rm \odot}),
\end{equation}
where $Z_{\rm cur}$ is the current metallicity.
When considering the temperature dependence, \cite{Sander2023} adjusted the mass loss of the He wind as follows:
\begin{equation}\label{}
    \centering
     \log \left(\frac{\dot{M}_{\rm SV2023}}{M_\odot\,\rm yr^{-1}}\right) = \log \left(\frac{\dot{M}_{\rm SV2020}}{M_\odot\,\rm yr^{-1}}\right) - 6\, \log \left(\frac{T_{\rm eff,crit}}{\rm 141 kK}\right),
\end{equation}
where $T_{\rm eff,crit}$ represents the effective temperature at the critical ($\approx$ sonic) point, and the fixed value of 141 kK accurately represents the $T_{\rm eff,crit}$ \citep{Sander2023}. To incorporate both theoretical and empirical insights, we adopt a hybrid wind model combining the prescriptions of \citet{Sander2023} and \citet{Vink2017}, hereafter referred to as SV2023+.

In \texttt{MESA}, the standard mass-loss prescriptions are unified into the Dutch wind scheme, which employs the rates of \citet{Vink2001} for hot, hydrogen-rich stars, \citet{deJager} for cool stars, and \citet{Nugis2000} for hot stars that have lost their hydrogen envelopes. For He stars, the \citet{Nugis2000} mass-loss prescription is implemented in \texttt{MESA} as

\begin{equation}\label{}
\log \left(\frac{\dot{M}_{\rm NL2000}}{M_\odot\,\mathrm{yr}^{-1}}\right) = -11.00 + 1.29 \log L + 1.73 \log Y + 0.47 \log Z_{\rm cur},
\end{equation}
where $Z_{\rm cur}$ is the current metallicity and $Y$ is the helium abundance.

\begin{figure*}
    \centering  
    \includegraphics[width=0.99\textwidth]{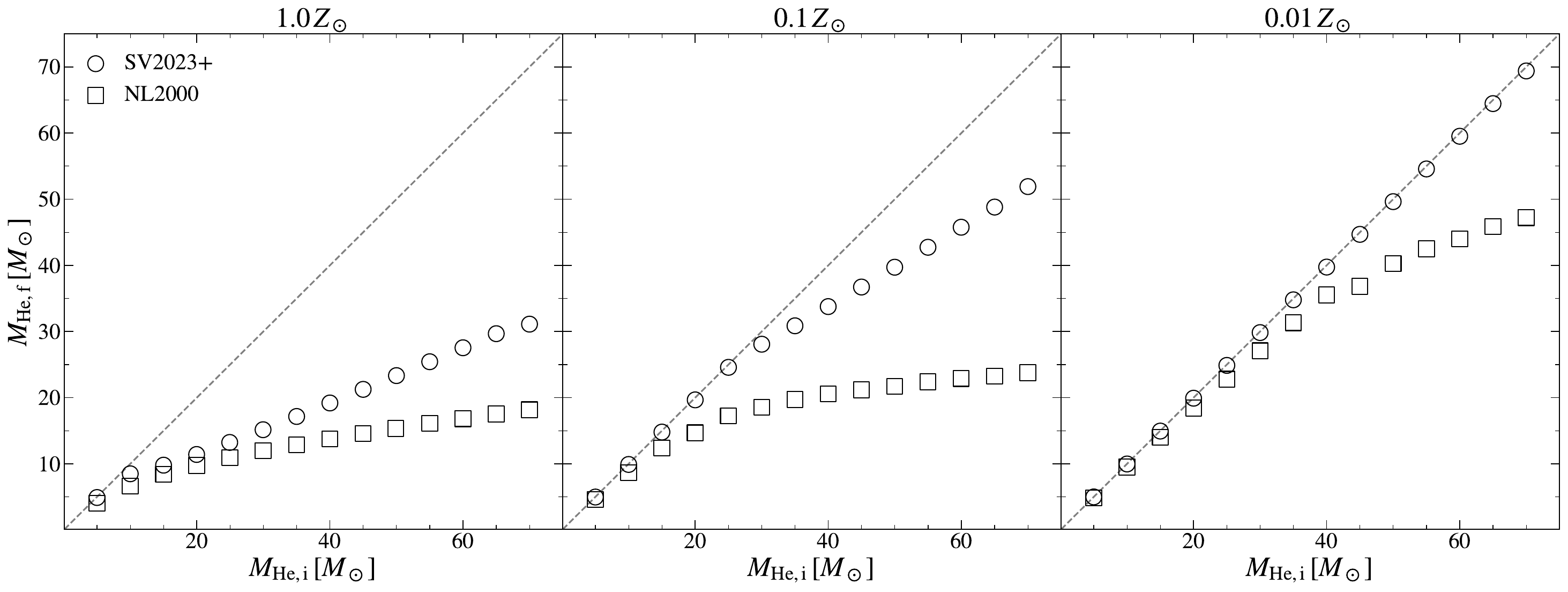}
    \caption{Final masses of He stars as a function of their initial masses with different wind prescriptions (\textit{left panel}: 1.0 $Z_{\odot}$; \textit{middle panel}: 0.1 $Z_{\odot}$; \textit{right panel}: 0.01 $Z_{\odot}$). Circle: SV2023+, square: NL2000. The dashed line indicates where the final mass is equal to the initial mass.}
    \label{Mhe_f}
\end{figure*}

Combining the two wind prescriptions, we evolve single He stars to examine the relation between their initial mass (He ZAMS) and final mass (at central carbon depletion). The initial He-star masses are chosen to span $5$–$70\,M_\odot$ at three different metallicites ($1.0\,Z_\odot$, $0.1\,Z_\odot$, and $0.01\,Z_\odot$). The results are shown in Figure~\ref{Mhe_f}.

At solar metallicity (left panel), He stars with initial masses $M_{\rm He,i}\gtrsim20M_\odot$ undergo substantial mass loss under the SV2023+ wind prescription, losing more than half of their initial mass. The NL2000 prescription yields a qualitatively similar correlation between initial and final mass but drives systematically stronger winds for $M_{\rm He,i} \gtrsim 20\,M_\odot$, resulting in even greater mass loss in this regime.

At low metallicity ($0.1\,Z_\odot$, middle panel), mass loss is generally weaker. Specifically, for He stars with $M_{\rm He,i} \lesssim 25\,M_\odot$, the SV2023+ prescription predicts negligible wind-driven mass loss. Notably, at $0.01\,Z_\odot$ (right panel), He stars show almost no mass loss across the entire mass range.

\section{Results}\label{sect3}

\subsection{Second-born BH spin magnitudes in BH--He binaries}
A total of 153 confident BBH events with a false-alarm rate below $1~\mathrm{yr}^{-1}$ have been reported in the GWTC-4.0 \citep{GWTC4_catlog}. Using the Hubble time as the maximum merger timescale from \citet{Peters1964}, we apply Kepler’s third law to estimate the maximum orbital period at birth for each coalescing binary system:
\begin{equation}\label{}
P_{\rm orb,\,max}\, [\mathrm{d}] = 
\frac{1}{24}\left( 
\frac{t_{\rm merger}}{9.829 \times 10^{6} \, {\rm yr}} 
\frac{M_{\rm BH_1} M_{\rm BH_2}}{M_{\rm total}^{1/3}}
\right)^{3/8},
\end{equation}
where $M_{\rm BH_1}$ and $M_{\rm BH_2}$ are the component BH masses, and $M_{\rm total}$ represents the total system mass, $t_{\rm merger}$ is expressed in years (assuming a circular orbit), and masses are in units of solar mass.

\begin{figure}[h]
     \centering
     \includegraphics[width=0.499\textwidth]{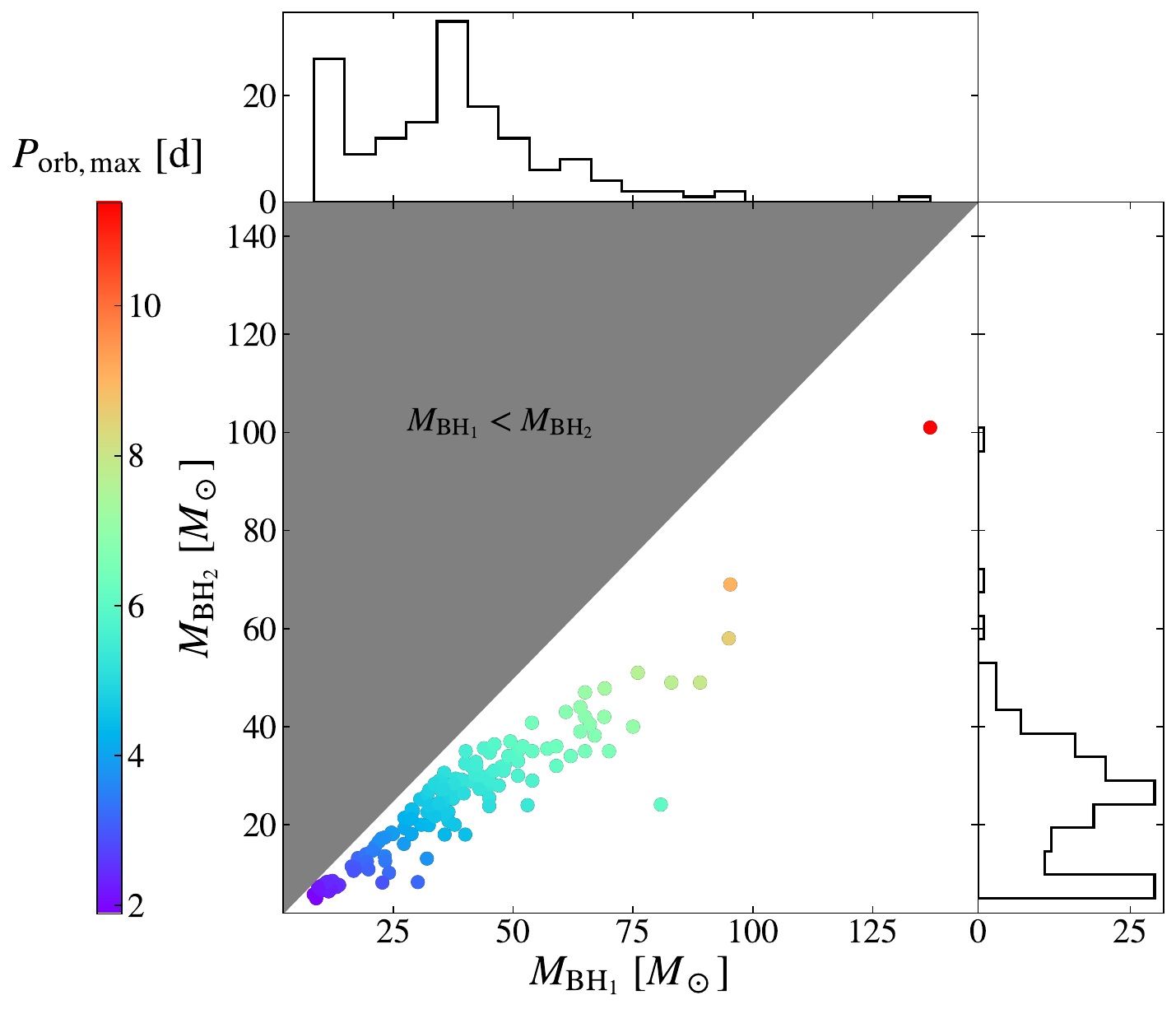}
     \caption{The color represents the maximum initial orbital period of the BBHs at birth, above which the systems would not merge within a Hubble time. Histograms of the primary component masses (\textit{upper panel}) and the secondary component masses (\textit{right panel}).}
     \label{p_max} 
\end{figure}

Figure~\ref{p_max} shows the distribution of the component masses and their corresponding maximum orbital periods ($\lesssim 11.0$ d) at birth. We find that the maximum orbital period increases with the component masses. This trend reflects the fact that more massive binaries emit gravitational waves more efficiently and therefore inspiral faster at a given separation. Consequently, they can be born in wider orbits and still merge within a Hubble time, whereas less massive binaries must start in tighter orbits to coalesce on the same timescale. Given that the immediate progenitor of a BBH is a close binary consisting of a BH and a He star, and that the He star undergoes substantial wind-driven mass loss, the orbital period of the BH--He system at birth is expected to be much shorter, since the orbit can otherwise be considerably widened by the wind mass loss. 

In the context of the isolated binary evolution, after the common envelope or stable mass transfer phase, the binary system—comprising a BH and a He star—continues to evolve in a close orbit. \citet{Detmers2008} were the first to investigate He stars in binaries with compact objects and found that He stars can not be tidally spun up to produce a collapsar at high metallicity due to the strong wind mass-loss rates adopted in their models. This effect is enhanced at low-metallicity environments because stellar winds are weaker, allowing He stars to retain angular momentum and be significantly tidally spun up \citep{Ma2023,Sen2025}. As demonstrated by \citet{Qin2018}, tidal interactions during the subsequent evolutionary phase are critical in determining the final spin of the newly formed BH. The efficiency of tidal spin-up may depend sensitively on the mass of the companion (i.e., the first-born BH). Moreover, the evolutionary stage of the He star at the onset of the BH–He star phase can affect the angular momentum budget of the BH progenitor, and hence the spin magnitude of the resulting BH—an effect that has not been previously explored. Using the SV2023+ wind prescription, we systematically investigate how the initial properties of the He star influence the final BH spin, including its initial rotation rate, mass, orbital period, and metallicity.

\subsubsection{BH spin on the its companion mass}
\begin{figure*}[h]
    \centering
    \begin{minipage}[t]{0.44\textwidth}
        \centering
        \includegraphics[width=\textwidth]{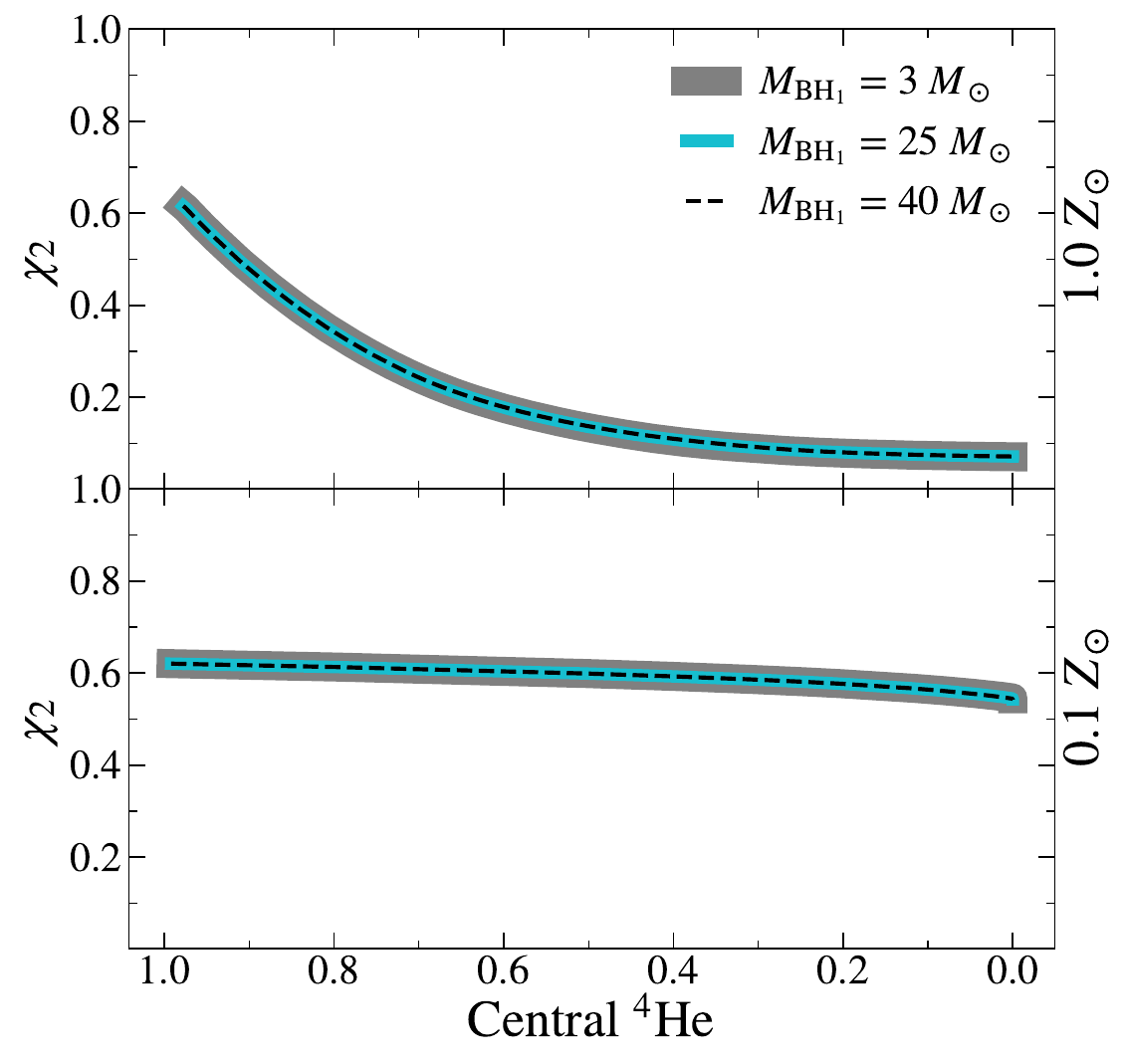}
    \end{minipage}
    \begin{minipage}[t]{0.45\textwidth}
        \centering   
        \includegraphics[width=\textwidth]{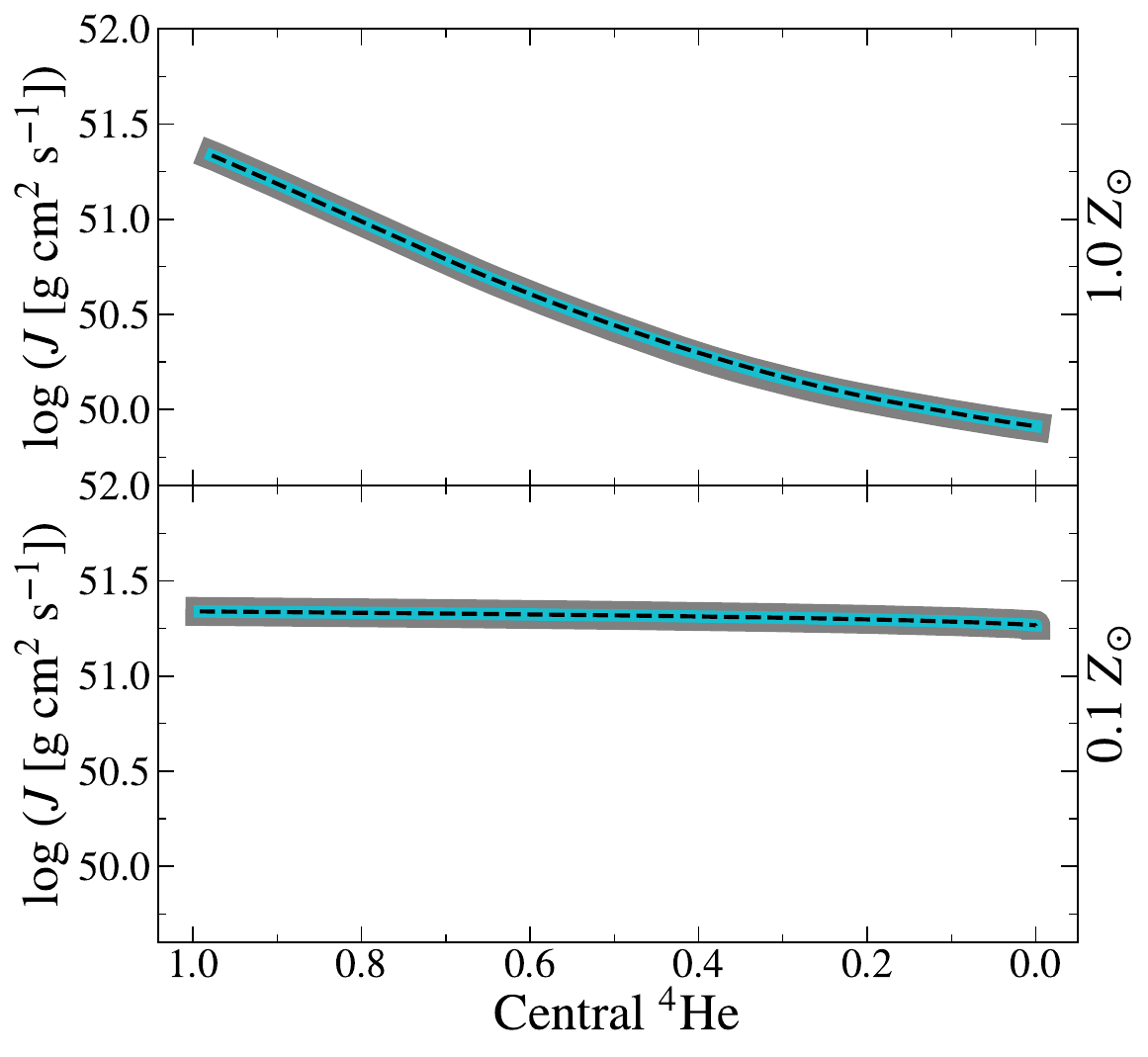}
    \end{minipage}
    \caption{The spin magnitude $\chi_2$ of the BH formed via direct core collapse of the He star (\textit{left panel}) and the corresponding total angular momentum of the progenitor star (\textit{right panel}) as a function of the central helium abundance. The binary system consists of a He star with an initial mass of 20 $M_{\odot}$ in a 1.0-day orbit, and a BH of varying mass (gray solid line: 3 $M_{\odot}$; cyan solid line: 25 $M_{\odot}$; black dashed line: 40 $M_{\odot}$). The results at 0.01 $Z_{\rm \odot}$ are nearly identical to those at 0.1 $Z_{\rm \odot}$, and are therefore not shown in the paper.}
    \label{diff_BH}
\end{figure*}

In general, a more massive BH companion typically shortens $T_{\rm sync}$ \cite[see Eq.(4) in][]{Qin2018}, leading to more efficient tidal spin-up of the He star. \cite{Qin2018} investigated the impact of the BH companion mass on the spin magnitude of the resulting BH, but considered only companion masses of 10 and 30 $M_\odot$. To investigate whether the BH mass significantly affects the final spin of the newly-formed BH, we model a 20 $M_\odot$ He ZAMS star paired with BH masses in a wide range of 3.0, 25, and 40 $M_\odot$ in an initial orbit of 1.0 d. For simplicity, we assume that the He star is initially synchronized with the orbit. Notably, as shown in the upper-left panel of Figure~\ref{diff_BH}, the BH spin magnitude $\chi_2$ remains largely unaffected by the BH companion mass throughout the evolution. As a comparison, we present in the bottom-left panel the results for He star models at a metallicity of Z = 0.1 $Z_\odot$, showing similar results. This indicates that the impact of varying BH mass on the final BH spin is negligible. Additionally, we also test the same binaries with a short initial orbit of 0.5 d, showing similar results as compared with a longer initial orbit (see Figure~\ref{appendixA_3}). In the right panels, we also present the total angular momentum of the progenitor star corresponding to each model.

\subsubsection{BH spin on the evolutionary stage of its progenitor He star}
Tidal interactions between a BH and a He star can affect the angular momentum of the progenitor, potentially influencing the spin of the resulting BH. However, it remains unclear whether the He star has significantly evolved by the time the system forms. To assess whether the spin of the resulting BH can be significantly altered under these conditions, we perform detailed binary evolution modeling for a system comprising a 20 $M_\odot$ BH and a 20 $M_\odot$ He star with different metallicities (1.0 $Z_{\odot}$, 0.1 $Z_{\odot}$, and 0.01 $Z_{\odot}$). For all cases, we adopt an initial orbital period of 1.0 and 0.5 d. We examine three different He star models as follows:

\begin{itemize}
\item{Zero-age helium main sequence star(He ZAMS})
\item{He star with 10\% of its central helium burned}
\item{He star with 30\% of its central helium burned}
\end{itemize}

\begin{figure*}[h]
     \centering     
     \includegraphics[width=0.4\textwidth]{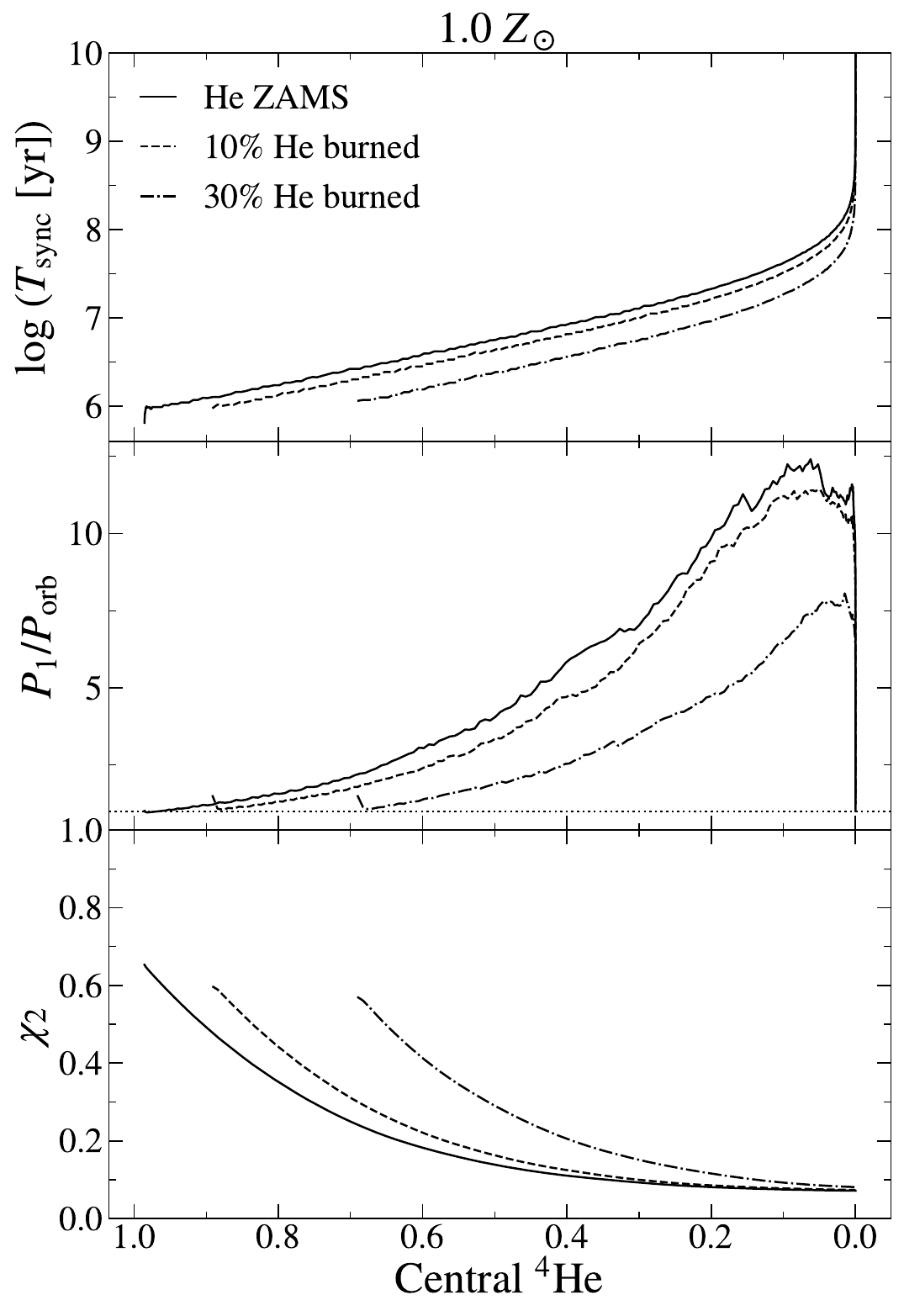}
     \includegraphics[width=0.41\textwidth]{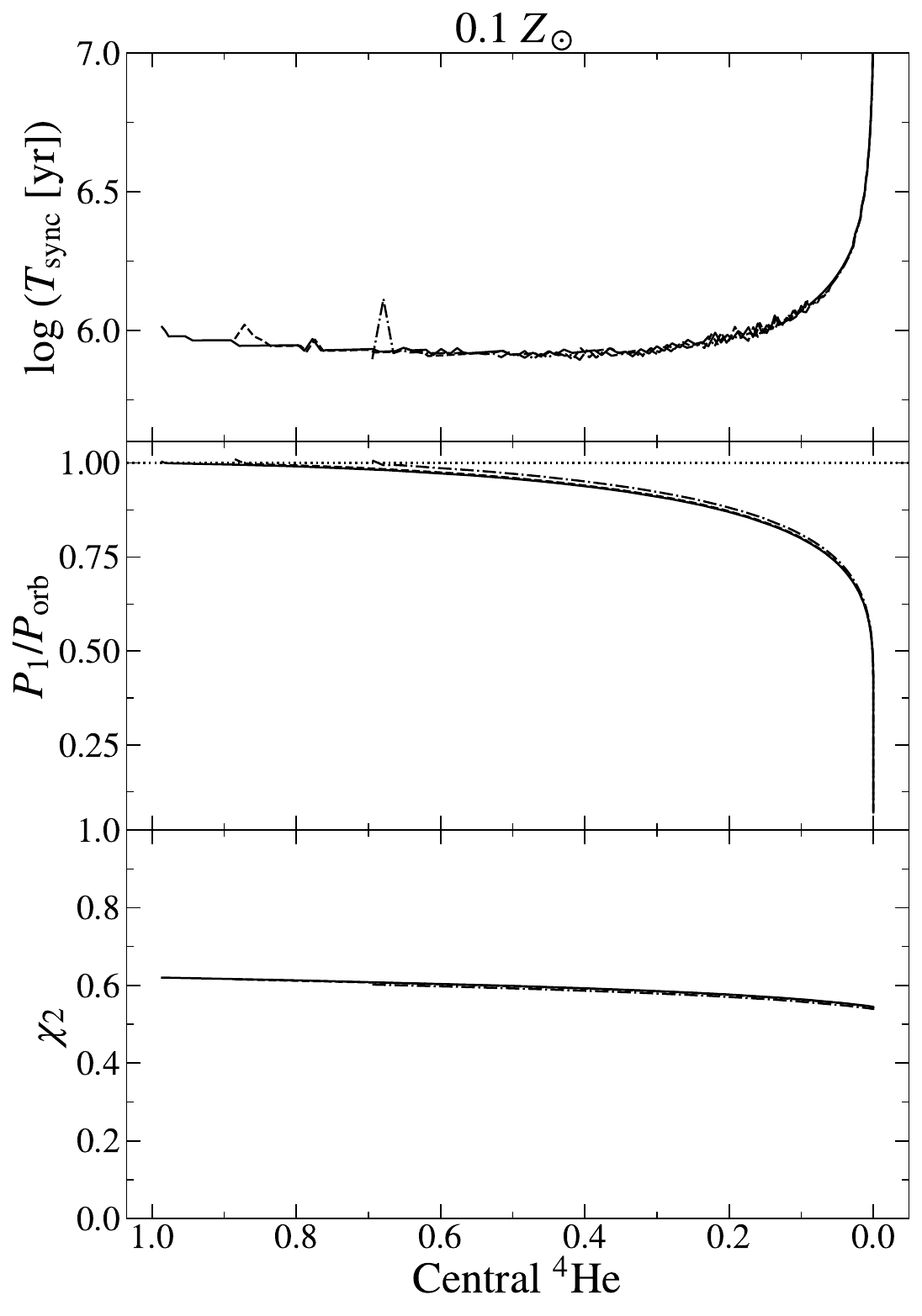}
     \caption{The synchronization timescale ($T_{\rm sync}$; top panel), the ratio of the He-star spin period to the orbital period ($P_1/P_{\rm orb}$; middle panel), and the BH spin magnitude ($\chi_2$; bottom panel) are shown as functions of the central helium abundance. We consider three He-star models: He ZAMS (solid line), a He star with 10\% central helium depleted (dashed line), and a He star with 30\% central helium depleted (dash-dotted line). All models assume an initial orbital period of 1.0 day. The first column shows the results at $1.0\,Z_{\odot}$, while the second column shows those at $0.1\,Z_{\odot}$. The results at $0.01\,Z_{\odot}$ are nearly identical to the $0.1\,Z_{\odot}$ case and are therefore not shown in the paper.}
     \label{Tsync_chi2} 
\end{figure*}

The He star is assumed to be initially synchronized with the orbit, and we will explore the implications of this assumption in a later section. We also assume that the progenitor star, at the end of its evolution, collapses directly into a BH without losing mass or angular momentum. The synchronization timescale ($T_{\rm sync}$) represents the time required for tidal interactions to bring the star’s spin into sync with the orbital period. We refer readers of interest to the updated formulation provided in \cite{Sciarini2024} and its implementation in \cite{Qin2024_gap}. We first show the results in the left column of Figure~\ref{Tsync_chi2}. In the upper panel, we show that $T_{\rm sync}$ evolves as the He star approaches central carbon depletion. During the core-helium burning phase, $T_{\rm sync}$ increases slightly due to orbital widening caused by wind-driven mass loss. After central helium depletion, tidal effects become negligible as the synchronization timescale $T_{\rm sync}$ increases rapidly. This is primarily due to the recession of the convective core, which leads to a significant reduction in the tidal torque coefficient $E_2$ \cite[see Equation (9) in][]{Qin2018}. The orbit gradually widens due to the strong wind from the He star at solar metallicity, which spins down of the He star and increases the ratio of spin period to orbital period ($P_1/P_{\rm orb}$) (see the middle panel). In the bottom panel, the spin magnitude $\chi_2$ of the resulting BH also decreases but converges to the same value for all three He-star models. The right column shows the corresponding evolution for the same He star at a lower metallicity of 0.1 $Z_{\odot}$. In this case, the wind is weaker and the orbit does not widen. As a result, the He star remains almost tidally locked. In the late evolutionary phase the star contracts \cite[see Figure 1 in][]{Qin2023}, spinning faster as angular momentum is conserved. Tidal torques then transfer angular momentum from the He star back into the orbit. Again, the final BH spin magnitude $\chi_2$ converges to nearly the same value across all three models.

The results above indicate that at high metallicity, strong stellar winds primarily govern the angular momentum evolution of He stars, whereas at low metallicity, weaker winds allow tidal interactions to play the dominant role. In both regimes, whether the He star is already evolved at the onset of the BH–He star phase has only a minimal effect on the spin magnitude of the resulting BH. For systems with even shorter initial orbital periods (i.e., 0.5 d), we observe a similar trend (see Figure~\ref{appendixA_1}).

\subsubsection{BH spin on the initial rotation of the He star}
\begin{figure*}[h]
    \centering
    \includegraphics[width=0.43\textwidth]{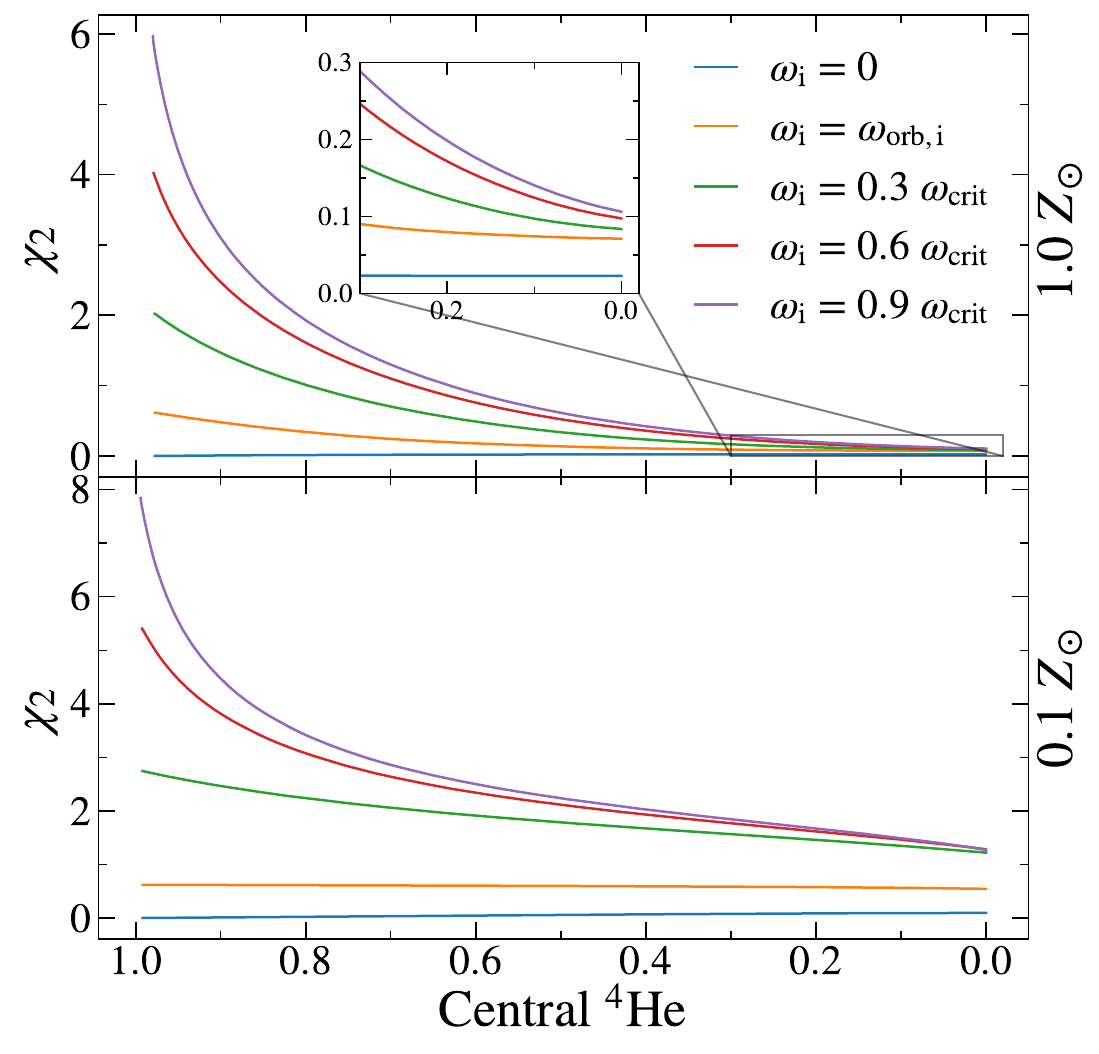}
    \includegraphics[width=0.44\textwidth]{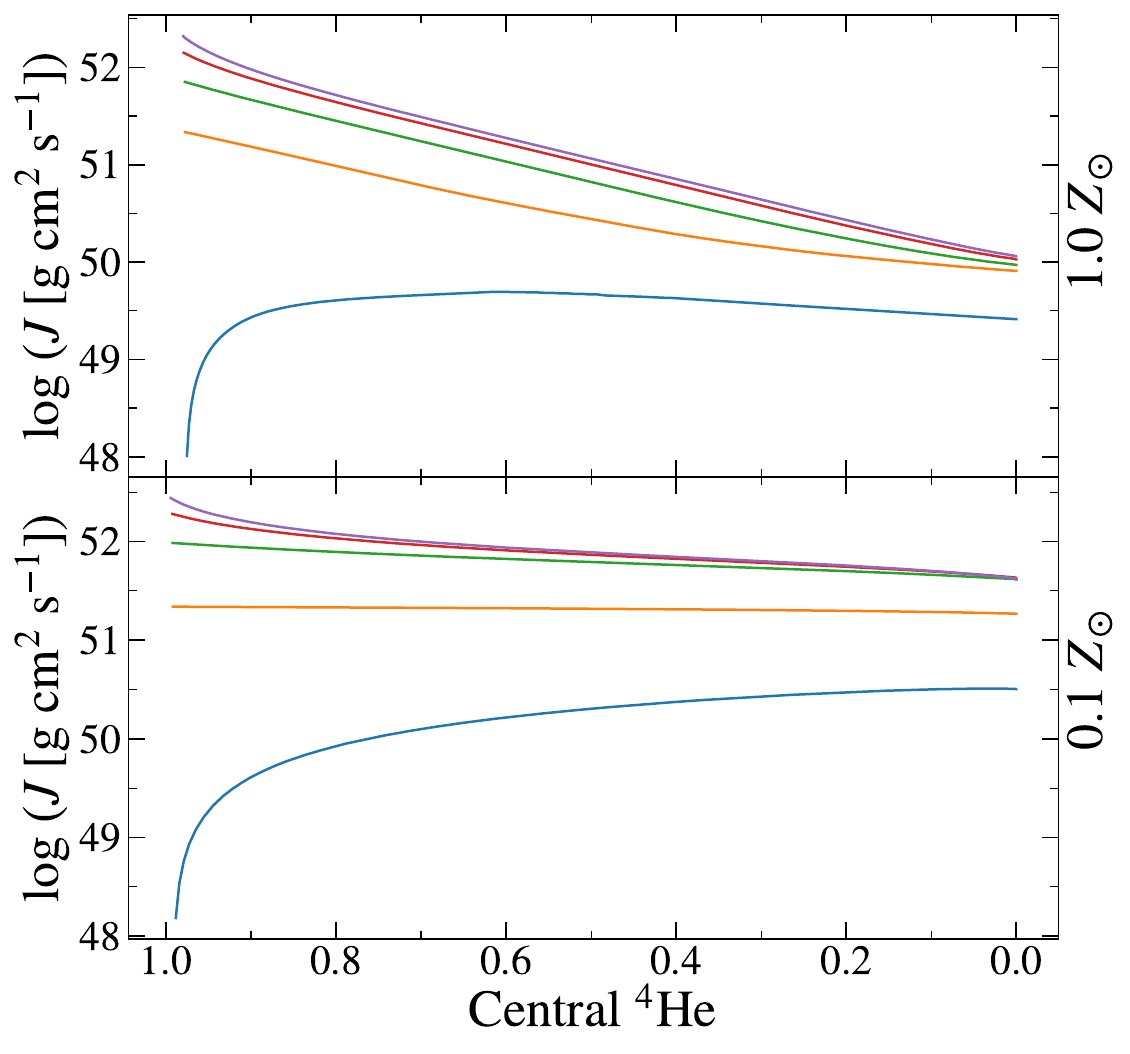}    
    \caption{The spin magnitude $\chi_2$ of the BH formed through the direct core collapse of the He star (\textit{left panel)} and the corresponding total angular momentum of the progenitor star (\textit{right panel}) as a function of the central helium abundance. The binary system consists of a 20 $M_{\odot}$ BH and a 20 $M_{\odot}$ He star with different initial rotation rates (blue: $\omega_{\rm i} = 0$; orange: initially synchronized with the orbit; green: $\omega_{\rm i} = 0.3\,\omega_{\rm crit}$; red: $\omega_{\rm i} = 0.6\,\omega_{\rm crit}$; purple: $\omega_{\rm i} = 0.9\,\omega_{\rm crit}$). All models assume an initial orbital period of 1.0 d, with the 0.5 d case provided for comparison in Figure~\ref{appendixA_2}. Results for He stars at 0.01 $Z_{\rm \odot}$ are nearly identical to those at 0.1 $Z_{\rm \odot}$ and are therefore not shown in the paper.}
    \label{diff_w}
\end{figure*}

The spin state of He stars at the formation of close BH–He star binaries remains poorly constrained. Although \citet{Qin2018} explored the role of He-star rotation in setting the spin of the resulting BH, their study considered only three initial rotation rates. Here, we conduct a systematic exploration of the dependence of the secondary BH spin magnitude on the initial rotation of the He star. To this end, we perform detailed binary evolution simulations for a system consisting of a 20 $M_\odot$ BH and a 20 $M_\odot$ He star in an initial orbit of 1.0 d with different initial metallicities (1.0 $Z_\odot$ and 0.1 $Z_\odot$), considering a range of initial He star spins: $\omega_i$ = 0, $\omega_i$ = 0.3 $\omega_{\rm crit}$, $\omega_i$ = 0.6 $\omega_{\rm crit}$, $\omega_i$ = 0.9 $\omega_{\rm crit}$, and synchronized with the orbit. In the upper-left panel of Figure~\ref{diff_w}, higher initial rotation rates (0.3–0.9 $\omega_{\rm crit}$) produce larger BH spin magnitudes at early times, but these values converge to $\sim$0.1 as the He star evolves. The synchronized model yields a slightly lower final spin ($\chi_2 \sim 0.07$), while the initially non-rotating model shows only negligible tidal spin-up. This demonstrates that, at high metallicity, the stellar wind dominates the angular-momentum evolution of He stars, causing the initial rotation to have only a minor impact on the final BH spin. In the bottom-left panel, at a lower metallicity, weaker wind mass loss allows the He star to retain more angular momentum. Since tides are also relatively inefficient, the initial rotation begins to play a much more significant role in determining the final BH spin magnitude. Therefore, the initial rotation rate of the He star tends to play a role at lower metallicities, especially when tides are inefficient. As a comparison, we run similar binary models at initially short orbits, showing that when tidal interactions are stronger, the differences in the final spin magnitude become smaller (see Figure~\ref{appendixA_2}). In the right panels, we also present the total angular momentum of the progenitor star corresponding to each model.

\subsubsection{BH spin on its progenitor He star's initial mass, metallicity, and orbital period.}
In addition to tidal interactions, wind mass loss from the He star is expected to play a crucial role in determining the spin of the resulting BH. More massive He stars and those with higher metallicities generally experience stronger stellar winds, leading to enhanced angular momentum loss. To quantify this effect, we perform binary evolution simulations of systems consisting of a 20 $M_{\rm \odot}$ BH and He stars with initial masses spanning 10--70 $M_{\rm \odot}$. We consider three initial metallicities for the He stars, i.e., 1.0 $Z_{\rm \odot}$, 0.1 $Z_{\rm \odot}$, and 0.01 $Z_{\rm \odot}$. For simplicity, we assume that the He star is initially synchronized with the binary orbit. The initial orbital periods are sampled uniformly in logarithmic space between 0.3 and 1.5 d.

\begin{figure*}[h]
  \centering
    \centering  
    \includegraphics[width=0.99\textwidth]{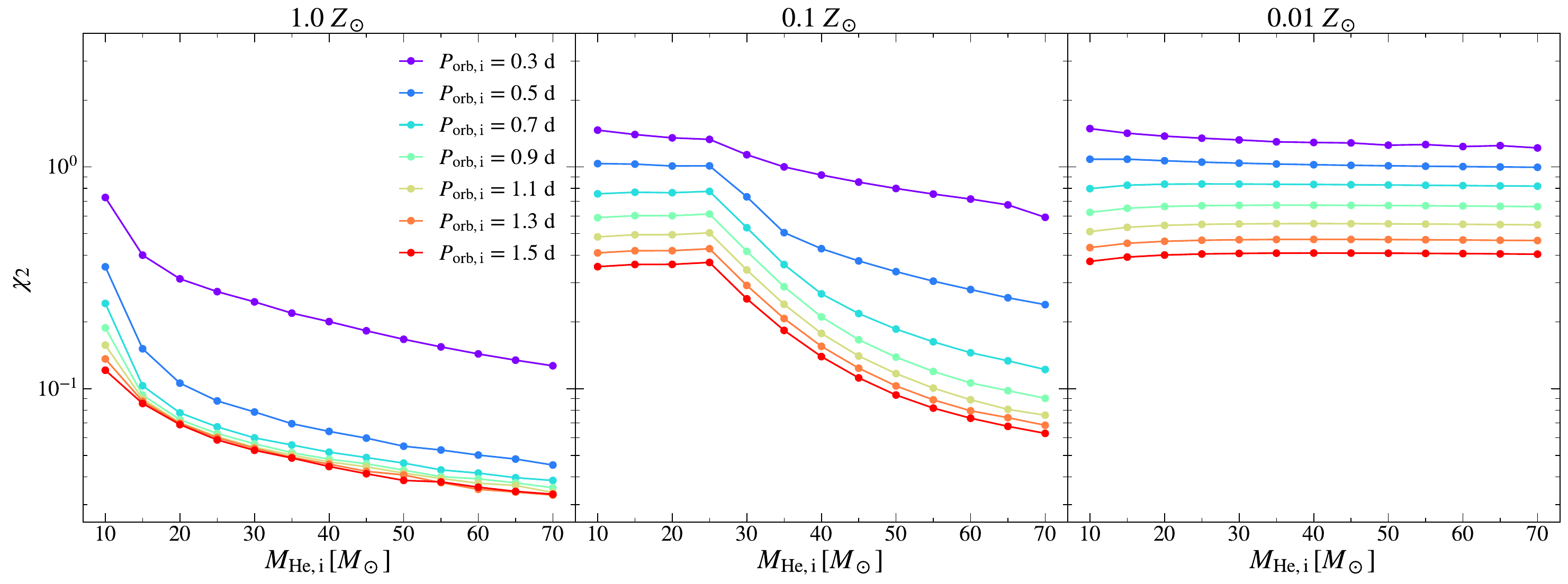}
    \caption{The spin magnitude $\chi_2$ of the resulting BH as a function of the initial He-star mass, sampled from 10 to 70 $M_{\odot}$ in a step of 5 $M_{\odot}$ (\textit{left panel}: 1.0 $Z_{\odot}$; \textit{middle panel}: 0.1 $Z_{\odot}$; \textit{right panel}: 0.01 $Z_{\odot}$). The calculation assumes angular-momentum conservation during direct core collapse at the time of central carbon depletion. Different colours denote different initial orbital periods.}
    \label{he_1_chi2}
\end{figure*}

Our results are presented in Figure~\ref{he_1_chi2}. We first examine the He stars at solar metallicity (left panel). For a fixed initial orbital period, the spin magnitude $\chi_2$ of the BH formed at central carbon depletion decreases with increasing initial He-star mass. This trend reflects the stronger stellar winds of more massive He stars, which more efficiently remove angular momentum. A similar dependence is observed with orbital period: wider orbits weaken tidal synchronization and reduce tidal spin-up, leading to lower angular momentum retained within the progenitor. These results indicate that although tidal interactions can efficiently spin up He stars in close binaries, wind-driven mass loss is the primary factor regulating the final BH spin magnitude at solar metallicity.

At lower metallicities, He stars are expected to retain more angular momentum owing to weaker winds. Consistent with this expectation, a similar mass-dependent decline in $\chi_2$ is observed at subsolar metallicity (middle panel) only for initial He-star masses $\gtrsim 25\,M_{\odot}$. For lower-mass He stars, the BH spin magnitude shows little sensitivity to the initial He-star mass. An analogous behavior is found at $0.01\,Z_{\odot}$ (right panel), where stellar winds are negligible (see Figure~\ref{Mhe_f}). In this regime, the final BH spin is largely set by the initial rotation of the He-star progenitor rather than by wind-driven angular momentum loss.

\subsection{Angular momentum transport efficiency on the resulting BH spin magnitude}
The Tayler–Spruit (TS) dynamo \citep{Spruit1999, Spruit2002}, driven by differential rotation in radiative layers, is considered a key mechanism for efficiently transporting angular momentum between a stellar core and its radiative envelope. Such transport plays a critical role in setting the spin magnitude of the BH formed from a hydrogen-rich massive progenitor star \citep{Qin2019}, although tidal interactions can also be important. \cite{Qin2023} further examined the TS dynamo in massive He stars, demonstrating that efficient angular momentum transport tends to enforce quasi–solid-body rotation.

\begin{figure}[h]
  \centering
    \centering  
    \includegraphics[width=0.45\textwidth]{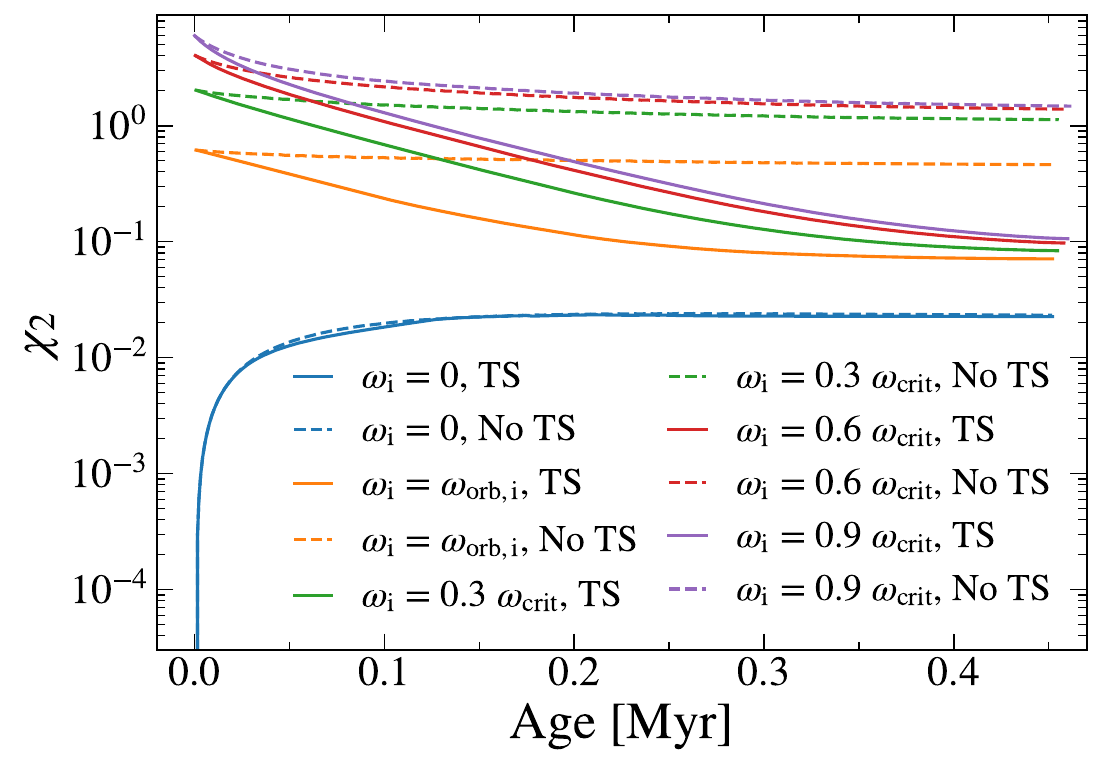}
    \includegraphics[width=0.44\textwidth]{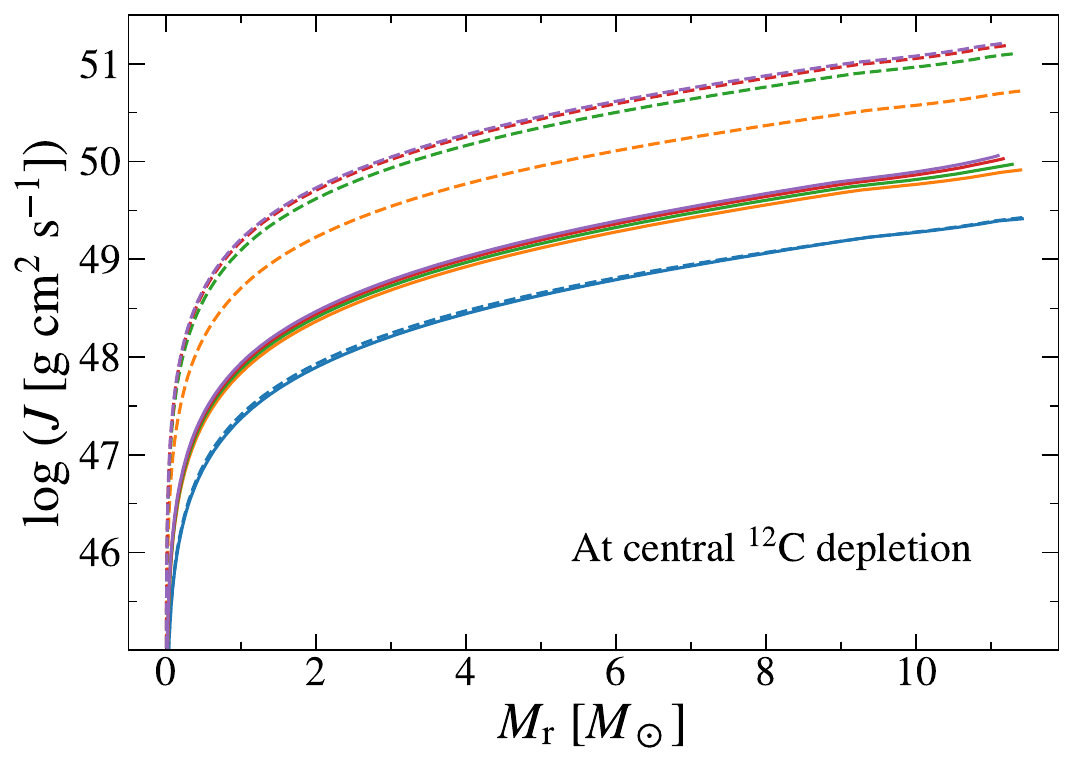}
    \caption{\textit{Upper panel:} same as Figure~\ref{diff_BH}, but for a binary with a $20\,M_\odot$ BH companion. Solid (dashed) lines denote models with (without) the TS dynamo. Lines with different colors correspond to various He star’s initial rotation rates (blue: $\omega_{\rm i}$ = 0; orange: initially synchronized with the orbit; green: $\omega_{\rm i} = 0.3\, \omega_{\rm crit}$; red: $\omega_{\rm i} = 0.6\, \omega_{\rm crit}$; purple: $\omega_{\rm i} = 0.9\, \omega_{\rm crit}$). \textit{Lower panel:} total internal angular momentum as a function of mass coordinate within the corresponding models at carbon depletion.}
    \label{chi2_j}
\end{figure}

The upper panel of Figure~\ref{chi2_j} shows the spin magnitude of BHs formed via the direct collapse of He stars at various evolutionary stages, for systems with an initial orbital period of $P_{\rm orb,i} = 1.0\,\mathrm{d}$ and a $20\,M_\odot$ BH companion. Different initial rotation rates are assumed for the He stars. The results indicate that the TS dynamo yields low-spin BHs (we assume the BH forms after the central carbon depletion), even when the progenitor He star begins with a high initial rotation rate. In the lower panel, we show that the total angular momentum retained in He stars at central carbon depletion differs by roughly an order of magnitude between models with and without the TS dynamo. Notably, the difference in spin magnitude and internal total angular momentum between models with and without the TS dynamo becomes negligible when He stars are assumed to be initially non-rotating, consistent with the findings of \cite{Bavera2020}. In the following, we focus on models incorporating efficient angular momentum transport (i.e., TS dynamo) within He stars to explore the correlation between mass ratio $q$ and inspiral effective spin $\chi_{\rm eff}$.

\subsection{Correlation between mass ratio $q$ and inspiral effective spin $\chi_{\rm eff}$}
The effective inspiral spin, $\chi_{\rm eff}$, a key diagnostic for distinguishing among the formation channels of merging BBH systems \citep[e.g.,][]{Abbott2016,Farr2017,Farr2018,Roulet2021}, is defined as
\begin{equation}
    \chi_{\rm eff} = \frac{M_{\rm BH_1} \overrightarrow{\chi_1} + M_{\rm BH_2} \overrightarrow{\chi_2}} {M_{\rm BH_1} + M_{\rm BH_2}} \hat{L},
\end{equation}
where $M_{\rm BH_1}$ and $M_{\rm BH_2}$ are the component BH masses, $\chi_1$ and $\chi_2$ are the corresponding dimensionless spin parameters, and $\hat{L}$ is the direction of the orbital angular momentum.

Given that the first-born BH typically has negligible spin \citep{Qin2018,Fuller2019}, the expression, under the assumption that $\chi_2$ is aligned with the orbital angular momentum, reduces to
\begin{equation}
    \chi_{\rm eff} =  \frac{q}{1 + q} \chi_2,
\end{equation}
where $q = M_{\rm BH_2}/M_{\rm BH_1}$ is the mass ratio of the binary system (hereafter, $M_{\rm BH_1}$ and $M_{\rm BH_2}$ denote the more massive and less massive BH, respectively).

As demonstrated earlier, the spin magnitude of the second-born BH is largely unaffected by the He star’s evolutionary stage at the onset of tidal interactions or the mass of its BH companion. However, mass loss due to stellar winds plays a dominant role, with more massive He star progenitors leading to lower-spin BHs. These findings underscore the efficiency of tidal synchronization in close BH--He binaries while highlighting the crucial impact of stellar winds on angular momentum evolution.

For simplicity, we assume that the He stars are initially synchronized with the orbit. Each He star evolves to carbon depletion and subsequently undergoes direct collapse to form a BH without mass or angular-momentum loss. The resulting spin magnitude should therefore be regarded as upper limits on the predicted BH spin. By extending the initial parameter space of the BH–He models, we derive a fitting formula for the spin of the second-born BH $\chi_2$ (see Appendix~\ref{appenB}), as a function of the initial He-star mass $M_{\rm He,i}$, orbital period $P_{\rm orb,i}$, and $Z_{\rm i}$.

To predict the correlation between $q$ and $\chi_{\rm eff}$, we first perform population synthesis calculations with \texttt{COMPAS} \citep[version v03.27.03;][]{Stevenson2017,Vigna2018,Neijssel2019,Riley2022,Disberg2025} to map the parameter distributions of BH–He star systems. Using the resulting parameter space, we then apply the fitting formula derived above to estimate the spin magnitude of the second-born BH. For the population synthesis, we adopt the fiducial model (see their Table 1) described in \cite{Hu2025}, with natal kick velocities drawn from a lognormal distribution following \cite{Disberg2025}. We adopt the same initial metallicity in both the population synthesis and the detailed binary evolution calculations, implicitly assuming that the metallicity does not evolve significantly during the transition from a main-sequence binary to a BH–He star system.

\begin{figure*}[h]
     \centering     
     \includegraphics[width=0.8\textwidth]{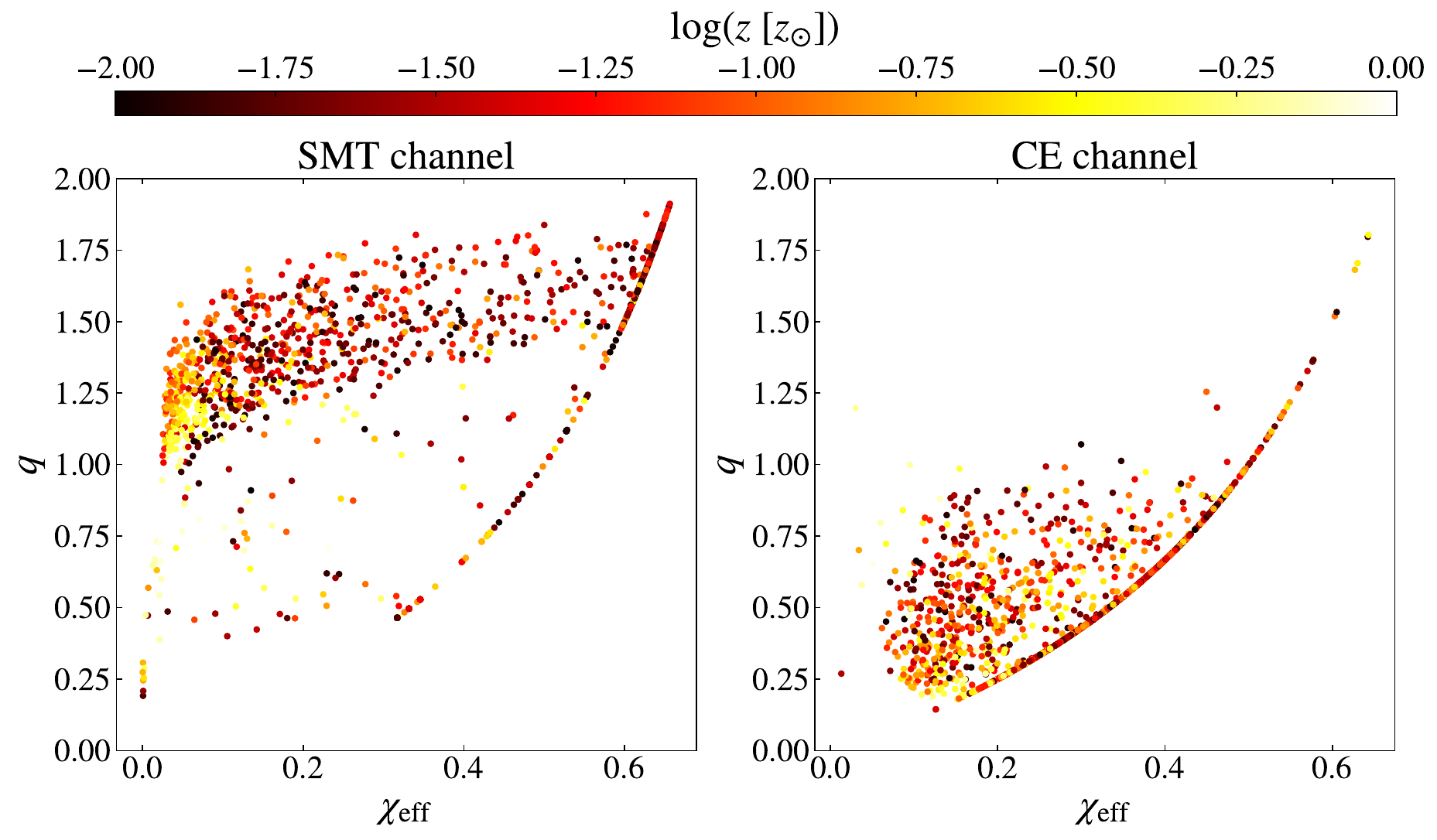}
     \caption{Predicted correlation between $q$ and $\chi_{\rm eff}$ for SMT (\textit{left panel}) and CE (\textit{right panel}) channel.}
     \label{q-xeff} 
\end{figure*}

After the formation of the first-born BH, the binary system undergoes either a phase of SMT or CE. Accordingly, the formation of BBHs can be divided into two evolutionary channels: the SMT channel and the CE channel. The resulting correlations between $q$ and $\chi_{\rm eff}$ for these two channels are shown in Figure~\ref{q-xeff}. For the SMT channel (left panel), we find that 85.8\% of BBHs have mass ratios $q > 1$. This arises because mass transfer from the initially more massive star to its companion during the first mass-transfer phase can reverse the mass ratio. In systems that undergo mass-ratio reversal, the accretor becomes more massive and then evolves rapidly to form the first-born BH. The accreted material carries angular momentum \citep{deMink2013}, which can spin up the accretor and potentially produce a rapidly rotating first-born BH. In our model, however, we assume that the first-born BH has negligible spin, which may lead to an underestimation. This assumption does not affect the CE channel, for which 2.8\% of BBHs also exhibit mass-ratio reversal (right panel). In addition, a large fraction of systems in the CE channel are found to have $\chi_{\rm eff} < 0.5$, consistent with theoretical expectations for BBH formation through CE evolution channel \citep{Qin2022}. Notably, no clear correlation between $q$ and $\chi_{\rm eff}$ is observed in either channel. This finding is based on the default assumptions of the fiducial model presented in \cite{Hu2025}. In future work, we will explore how different physical prescriptions adopted in population synthesis affect the relationship between the mass ratio $q$ and the effective spin parameter $\chi_{\rm eff}$.

\section{Conclusions and discussion}\label{sect4}
As the final evolutionary stage prior to BH formation, He stars play a pivotal role in setting the resulting BH mass through their mass-loss history. Here, we model the evolution of massive He stars using the physically motivated wind prescription of \citet{Sander&Vink2020}, together with the revised, temperature-dependent formulation of \citet{Sander2023}. In comparison with the standard Dutch wind scheme, commonly adopted in stellar evolution studies, the new prescription predicts significantly reduced mass loss.

Using the new wind prescription, we further examine key factors that may influence the spin magnitude of a BH inherited from its progenitor He star at different metallicities in close binaries. Our results show that the final spin is insensitive to both the evolutionary stage of the He star at the onset of tidal interactions and the mass of its companion. Moreover, the initial rotation of He stars has only a minor impact—particularly under strong tidal coupling—supporting the common assumption of initial orbital synchronization adopted in earlier studies \cite[e.g.,][]{Fragos2023}. For different initial orbital periods, we find that the spin magnitude systematically decreases with increasing BH mass, approximately following an exponential trend. 

Using detailed binary-evolution modeling, we derive a fitting formula for the spin magnitude of the resulting BH as a function of the properties of BH–He star systems at the time of their formation. This prescription can be readily implemented in population-synthesis studies to predict BH spin magnitudes. Using BH–He star populations generated by population-synthesis calculations with the fiducial model of \cite{Hu2025}, we investigate BBHs formed in both SMT and CE channels. In the SMT channel, the majority (85.8\%) of BBHs undergo mass-ratio reversal, whereas in the CE channel, only a small fraction (2.8\%) of BBHs exhibit mass-ratio reversal. Notably, we find no correlation between the mass ratio $q$ and the effective spin $\chi_{\rm eff}$ in either evolutionary channel. In future studies, we will investigate the impact of different physical prescriptions in population-synthesis models on the relationship between the mass ratio $q$ and the effective spin parameter $\chi_{\rm eff}$.

Predicting the properties of BBHs remains challenging owing to uncertainties in the physics of massive star evolution, both in isolation and in binaries \citep{Belczynski2022}. Recent models suggested that He stars may preserve a thin hydrogen layer on their surfaces after the common-envelope phase \citep{Nie2025}. Such an envelope can induce stellar expansion and trigger stable Case BB mass transfer onto the companion \citep{Tauris2015,Qin2024_casebb}, thereby causing further mass and angular momentum loss of the BH progenitor. Another major uncertainty concerns whether BHs receive natal kicks at birth. In the isolated binary scenario, BH spins are generally expected to align with the orbital angular momentum \citep{Kalogera2000,Farr2017}. BBH events with negative effective spins require at least one misaligned spin component, most likely induced by supernova kicks. However, strong natal kicks would be inconsistent with observations of Galactic BH binaries \citep[e.g.,][]{Mandel2016}. \citet{Tauris2022} showed that isolated binary evolution can still reproduce the observed BBH population if the BH spin axis is tossed during its formation in the core collapse of a massive star. Moreover, \citet{Baibhav2024} explored various physical mechanisms that can affect both the BH spin magnitudes and spin–orbit misalignments. We also note that the efficiency of angular momentum transport within massive He stars can substantially affect the spin magnitude of the resulting BH. Additionally, we do not account for accretion feedback during the core-collapse phase \citep{Batta2019}. Consequently, the spin magnitude of the resulting BH, obtained under the assumption of direct core collapse without mass or angular-momentum loss, should be regarded as an upper limit. Moreover, assuming direct core collapse at the time of central carbon depletion may not be appropriate for low-mass He stars, which can undergo significant expansion at later evolutionary stages and experience substantial mass loss through mass transfer onto their companions \citep{Wu2022}.

As the number of BBH events reported by the LVK Collaboration continues to grow, it has been suggested that the observed population is produced by a combination of multiple formation channels \cite[e.g.,][]{Zevin2021,Mandel2022,Cheng2023,Afroz2025,Colloms2025}. Among the various observables, the effective inspiral spin $\chi_{\rm eff}$ has been used to help distinguish the formation channels of some BBH events. For example, GW190517, which exhibits a high effective spin of $\chi_{\rm eff} = 0.52^{+0.19}_{-0.19}$ in GWTC-2 \citep{2021PhRvX..11b1053A}, has been suggested to originate from chemically homogeneous evolution \citep{Qin2022_che}. In contrast, GW191109, which has the most negative measured $\chi{\rm eff}$, has been associated with a dynamical formation channel \citep{Zhang2023}. More recently, the LVK Collaboration reported the heaviest BBH merger to date, GW231123, with a total mass of $190$–$265\,M_{\odot}$. This event may also originate from the chemically homogeneous evolution \citep{demink2025}; however, its formation channel remains under debate \citep{GW231123}.

\begin{acknowledgements}
 Y.Q. acknowledges support from the National Natural Science Foundation of China (grant Nos. 12473036 and 12573045) and Anhui Provincial Natural Science Foundation (grant No. 2308085MA29). This work was partially supported by the Jiangxi Provincial Natural Science Foundation (grant Nos. 20242BAB26012 and 20224ACB211001) and by Anhui Province Graduate Education Quality Engineering Project (grant No. 2024qyw/sysfkc012). G.M. has received funding from the European Research Council (ERC) under the European Union’s Horizon 2020 research and innovation program (grant agreement No 833925, project STAREX). H.F. Song is supported by the National Natural Science Foundation of China (grant Nos. 12173010 and 12573034). All figures are made with the free Python module Matplotlib \citep{Hunter2007}.
\end{acknowledgements}

\bibliography{ref}

\begin{appendix}
\section{Additional figures}
\begin{figure*}[h]
     \centering     
     \includegraphics[width=0.4\textwidth]{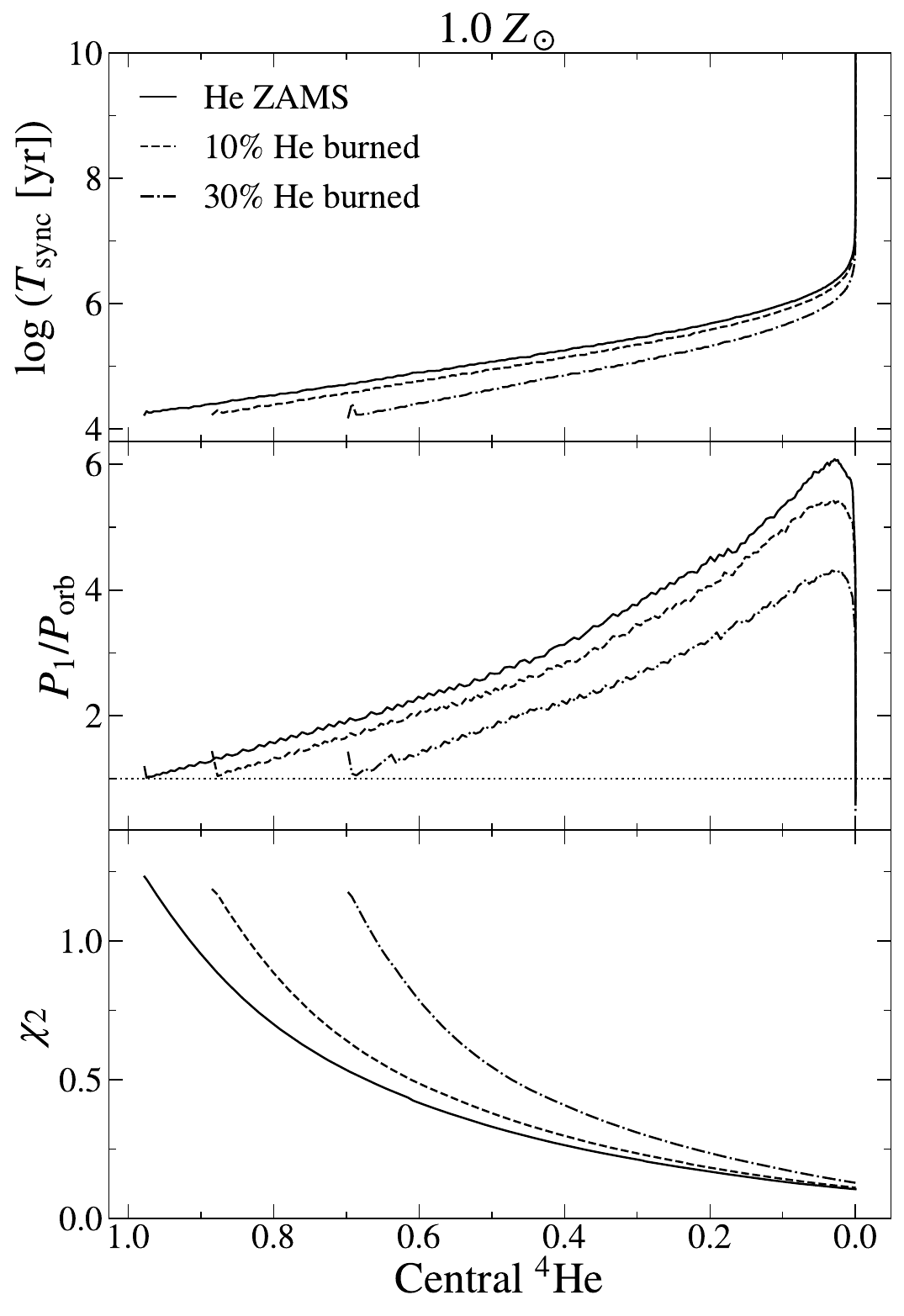}
     \includegraphics[width=0.41\textwidth]{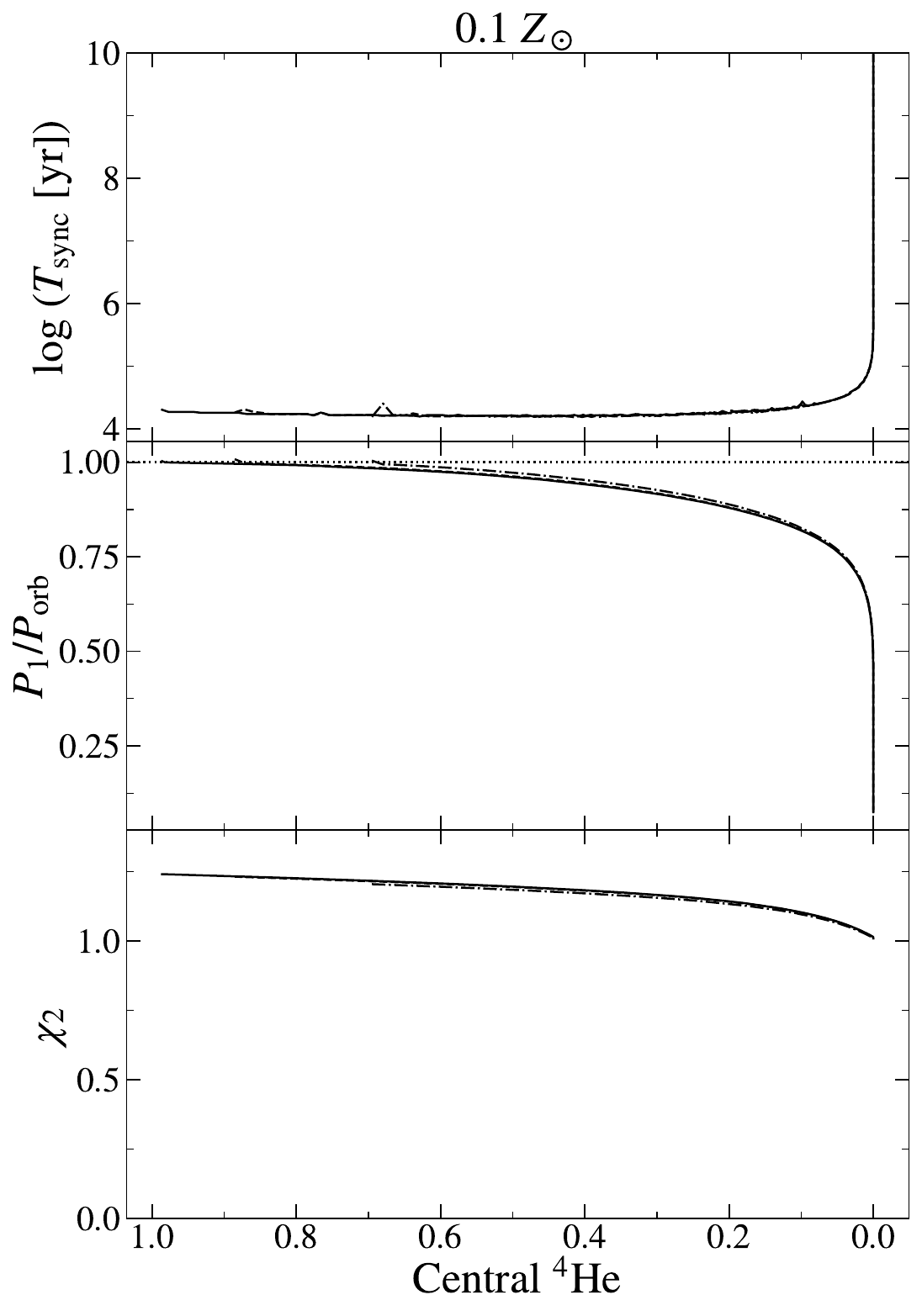}
     \caption{Similar to Fig.~\ref{Tsync_chi2}, but assuming an initial orbital period of 0.5 d.}
     \label{appendixA_1} 
\end{figure*}

\begin{figure}[h]
     \centering     
     \includegraphics[width=0.42\textwidth]{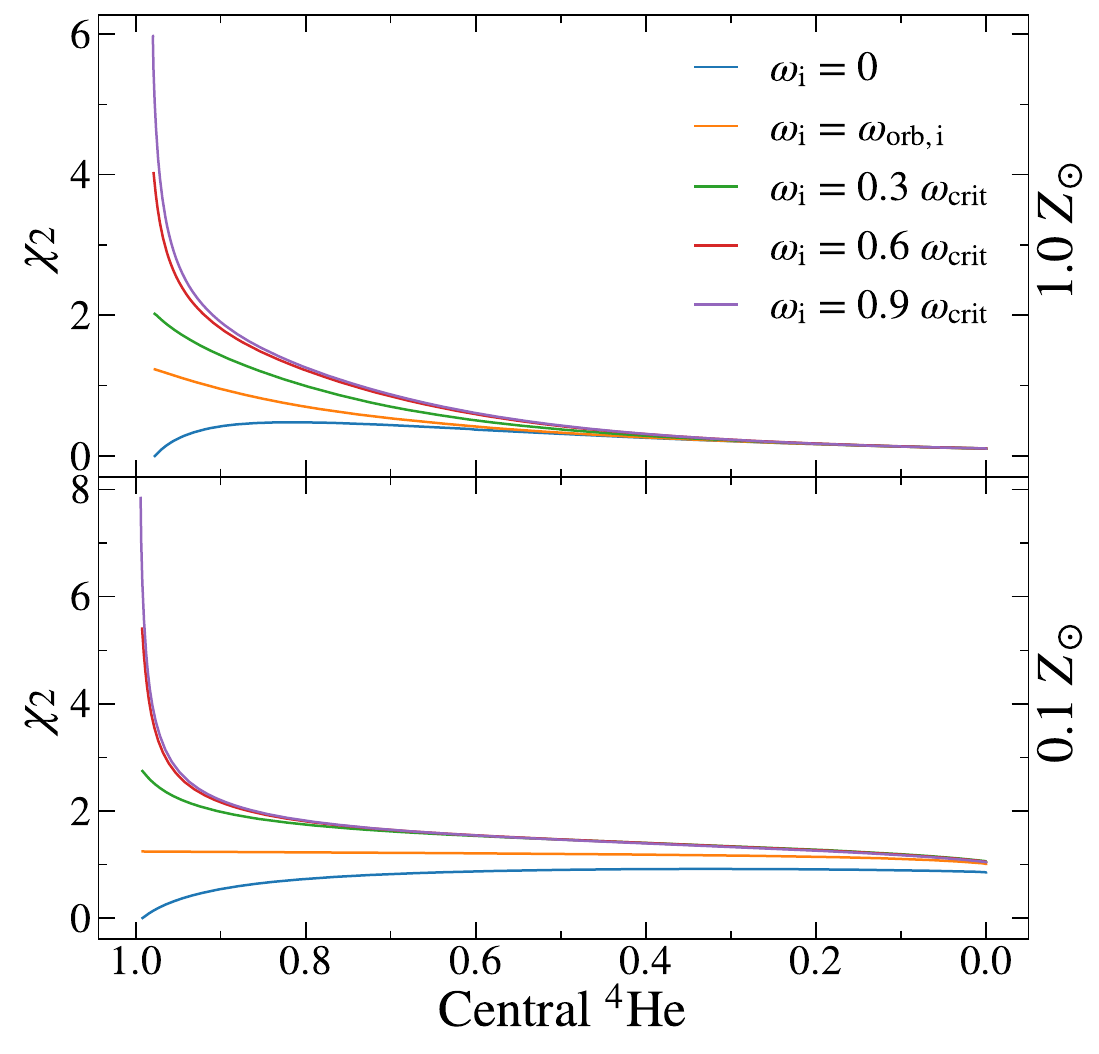}
     \caption{Similar to Fig.~\ref{diff_w}, but assuming an initial orbital period of 0.5 d.}
     \label{appendixA_2} 
\end{figure}

\begin{figure}[h]
     \centering     
     \includegraphics[width=0.44\textwidth]{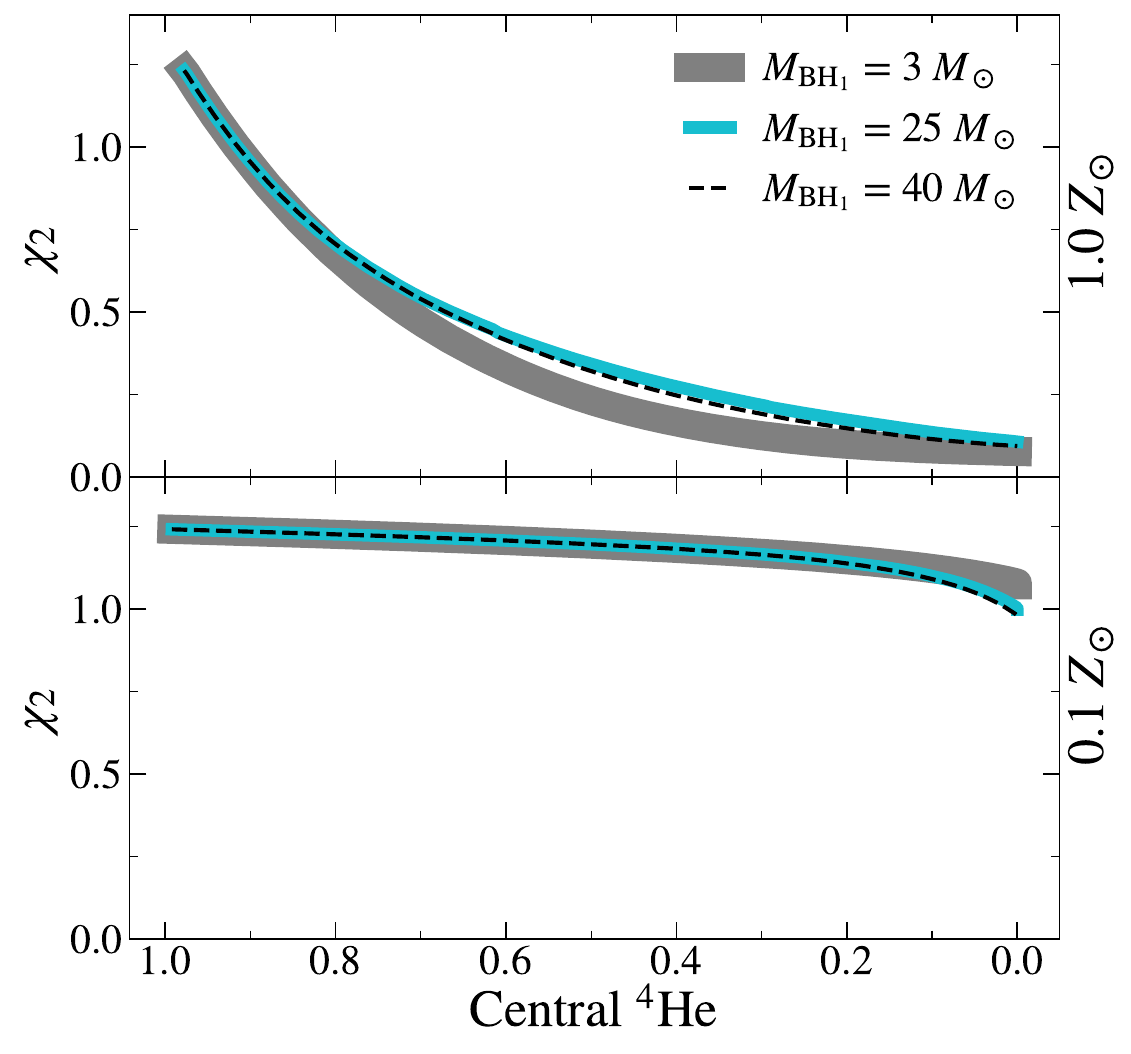}
     \caption{Similar to Fig.~\ref{diff_BH}, but assuming an initial orbital period of 0.5 d.}
     \label{appendixA_3} 
\end{figure}

\section{Fitting formula of the second-born BH spin $\chi_2$}\label{appenB}
Because the resulting BH spin $\chi_2$ is insensitive to the mass of the companion star, we construct \texttt{MESA} models of BH–He star binaries assuming a fixed companion point mass of $M_{\rm BH_1} = 20\,M_{\odot}$. We also assume that the He star is initially synchronized with the orbit. The initial He-star mass is sampled from 10 to 70 $M_{\rm \odot}$ in steps of 5 $M_{\rm \odot}$, while the initial orbital period ranges from 0.3 to 1.5 d with a step of 0.2 d. We consider 13 discrete values for the initial He-star metallicity, $Z_{\rm i} =$ 0.01, 0.03, 0.05, 0.08, 0.1, 0.3, 0.5, 0.8, 1.0, 1.3, 1.5, 1.8, and 2.0 $Z_{\rm \odot}$. In total, our final grid contains 1183 detailed binary models.

\begin{figure}[h]
     \centering
     \includegraphics[width=0.45\textwidth]{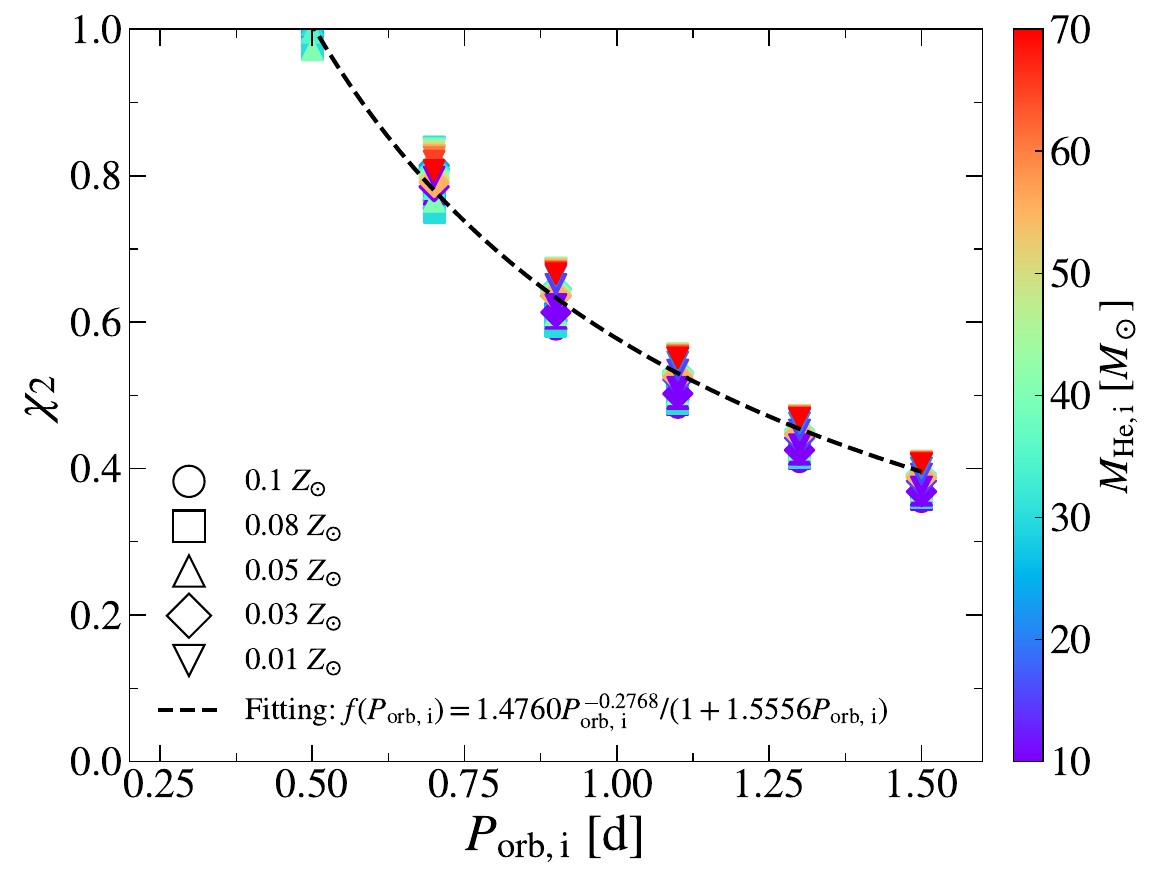}
     \caption{The spin magnitude $\chi_2$ as a function of the initial orbital period for the full range of He-star masses at low metallicities (($Z_{\rm i} \le 0.1\, Z_{\odot}$)). Different symbols denote different initial metallicities, i.e., circles: 0.1 $Z_\odot$, squares: 0.08 $Z_\odot$, upward triangles: 0.05 $Z_\odot$, diamonds: 0.03 $Z_\odot$, downward triangles: 0.01 $Z_\odot$. The dashed line shows the fitting formula for $\chi_2$ as a function of the initial orbital period across all He-star masses.}
     \label{low_z_fit} 
\end{figure}

\begin{figure}[h]
     \centering
     \includegraphics[width=0.45\textwidth]{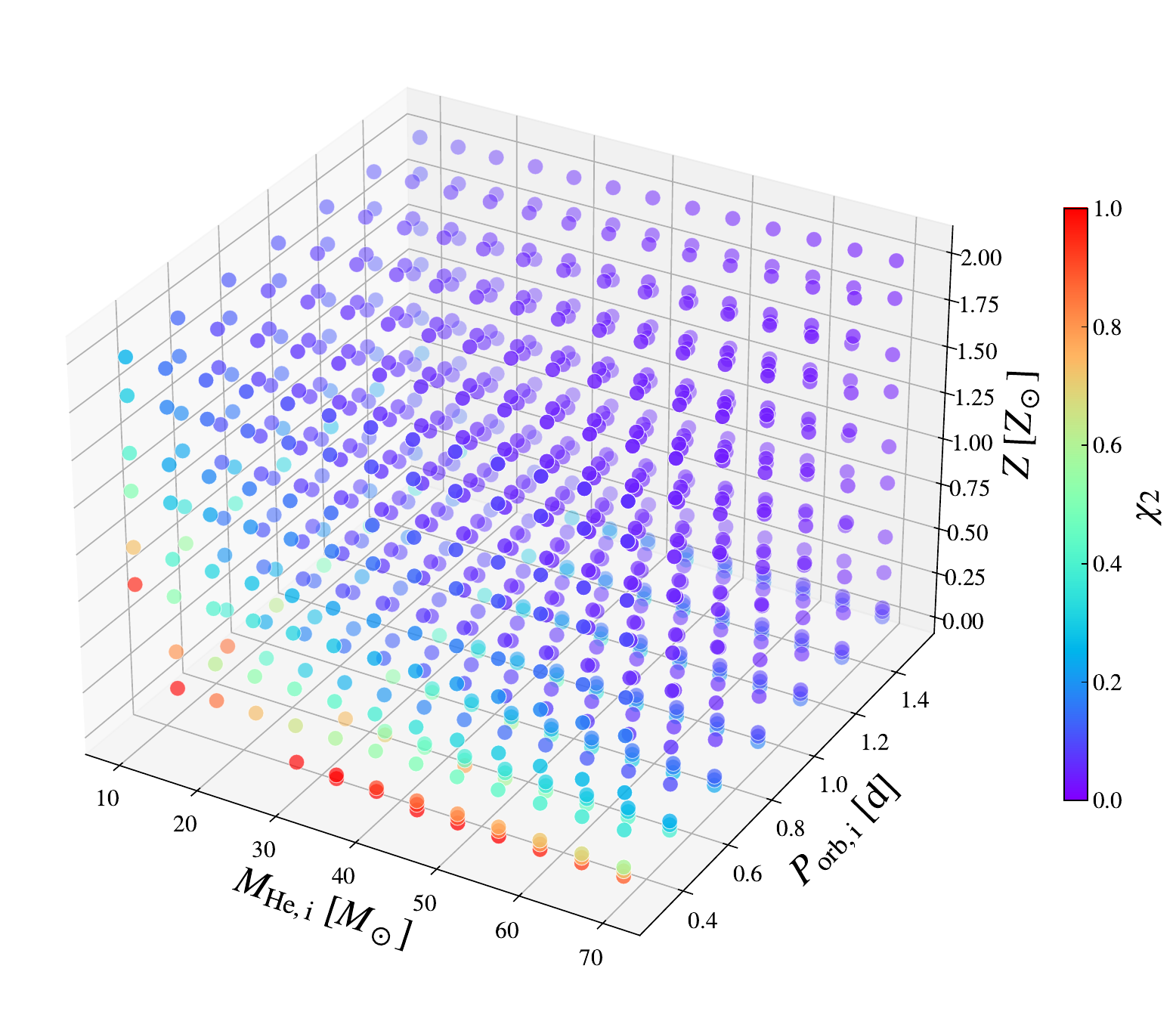}
     \caption{The spin magnitude (colorbar) of the resulting BH in close BH--He-star systems as a function of the initial He-star mass, orbital period, and metallicity ($Z_{\rm i} \ge 0.1\, Z_{\odot}$) at the end of the carbon depletion.}
     \label{3D_spin_True} 
\end{figure}

At low metallicity, He stars experience weak stellar winds. As a result, the spin magnitude of the resulting BH is primarily governed by tidal interactions, which depend sensitively on the orbital period between the BH progenitor and its companion. Figure~\ref{low_z_fit} shows that at low metallicities (i.e., $Z_{\rm i} \le 0.1 Z_\odot$), the BH spin decreases monotonically with increasing initial orbital period, independent of the He-star mass over the full mass range considered in this work. At high metallicity, He stars are expected to have strong wind mass loss. Therefore, the spin magnitude of the resulting BH is determined by the interplay between the tidal interaction and wind mass loss. For all the parameter space, we show in Figure~\ref{3D_spin_True} that the spin magnitude of the BH in close BH--He-star systems as a function of the initial He-star mass, orbital period, and metallicity at the end of the carbon depletion.

When the luminosity drops below a critical threshold (namely L$_0$, see Equation~\ref{L_L0}), it becomes insufficient to sustain optically thick winds, resulting in negligible mass loss for lower-mass He stars. For He stars across a range of metallicities, we derive a mass–luminosity relation, $L = 7626.85 M^{1.41}_{\rm He,i}$. Combining this relation with Equation~\ref{L_L0}, we express the initial He-star mass as a function of its initial metallicity as follows:
\begin{equation}
    M_{\rm He,\,i} = 6.9 \,Z_{\rm i}^{-0.62}.
    \label{}
\end{equation}

With detailed binaries models, we then derive a fitting formula for the spin magnitude of the resulting BH, $\chi_2$, as a function of $M_{\rm He,\,i}$, $P_{\rm orb,\,i}$, and $Z_{\rm i}$:
\begin{equation}
\chi_2 = 
\begin{cases} 
\min\{1,\,f(P_{\rm orb,\,i})\}, & M_{\rm He,\,i} < 6.9\, Z_{\rm i}^{-0.62} \\[0.5em]
\min\{1,\,f(M_{\rm He,\,i}, P_{\rm orb,\,i}, Z_{\rm i})\}, & \rm else \\[0.5em]
\end{cases}
\end{equation}
where
\begin{equation}
    f(P_{\rm orb,\, i})= \frac{1.4760P_{\rm orb,\, i}^{-0.2768}}{1 + 1.5556P_{\rm orb,\, i}},
    \label{eq:spin_model}
\end{equation}
\begin{equation}
    f(M_{\rm He, \,i}, P_{\rm orb,\, i}, Z_{\rm i})= \frac{4.1091M_{\rm He, \,i}^{-0.4666} P_{\rm orb,\, i}^{-0.7514}\exp({-0.4561Z_{\rm i})}}{1 + 0.7404M_{\rm He, \,i}P_{\rm orb,\, i}Z_{\rm i}} + 0.0327,
    \label{eq:spin_model}
\end{equation}
where $M_{\rm He, \,i}$, $P_{\rm orb,\, i}$, and $Z_{\rm i}$ are expressed units of solar mass, days, and solar metallicity, respectively.

We assess the fit quality using the coefficient of determination ($R^2$: the ratio of the regression sum of squares to the total sum of squares) and the root-mean-square error (RMSE: the standard deviation of the residuals between the fitted model and the data). Our best-fit model yields $R^2$ = 0.9602 and an RMSE of 0.0368, indicating a good agreement with the data.

We note that \cite{Bavera2021} provided an fitting fomula of the BH spin $\chi_2$ based on detailed binary models presented in \cite{Bavera2021B}. In their work, mass loss from He stars was treated using the stellar-wind prescription of \citet{Brott2011}. The updated approximation for the BH spin $\chi_2$ derived here instead adopts the SV2023+ stellar-wind prescription. In addition, we employ the revised tidal prescription proposed by \cite{Sciarini2024} (see also the comparisons presented in \cite{Qin2024_gap}.

\end{appendix}
\end{document}